\documentclass{article}

\usepackage{PRIMEarxiv}

\usepackage[square]{natbib}
\usepackage[utf8]{inputenc} 
\usepackage[T1]{fontenc}    
\usepackage{hyperref}       
\usepackage{url}            
\usepackage{booktabs}       
\usepackage{amsfonts}   
\usepackage{graphicx}
\usepackage{epstopdf}
\usepackage{float}
\usepackage{amsmath}
\usepackage{amssymb}
\usepackage{amsfonts}
\usepackage{nccmath}
\usepackage{mwe}
\usepackage[table]{xcolor}
\usepackage{subcaption}
\usepackage{booktabs} 
\usepackage{nicefrac}  
\usepackage{hyperref}
\usepackage{rotating}
\usepackage{nicefrac}       
\usepackage{graphicx}
\usepackage{booktabs}
\usepackage[toc,page]{appendix}
\usepackage{setspace}
\usepackage[linesnumbered,ruled]{algorithm2e}
\DeclareMathOperator*{\argmin}{arg\,min}

\newtheorem{proposition}{Proposition}[section]

\newtheorem{proof}{Proof:}[section]

\newcommand{\qed}{\tag*{$\blacksquare$}}

\graphicspath{{media/}}     

\title{Fast Computation of Branching Process Transition Probabilities via ADMM}

\author{
  Achal Awasthi \\
  Department of Biostatistics and Bioinformatics\\
  Duke University \\
  Durham, NC\\
  \texttt{{achal.awasthi}@duke.edu} \\
   \And
  Jason Xu\\
  Department of Statistical Science\\
  Duke University \\
  Durham, NC\\
  \texttt{jason.q.xu@duke.edu} \\
}

\begin{document}
\maketitle

\begin{abstract}
Branching processes are a class of continuous-time Markov chains (CTMCs) prevalent for modeling stochastic population dynamics in ecology, biology, epidemiology, and many other fields. The transient or finite-time behavior of these systems is fully characterized by their transition probabilities. However, computing them requires marginalizing over all paths between endpoint-conditioned values, which often poses a computational bottleneck. Leveraging recent results that connect generating function methods to a compressed sensing framework, we recast this task from the lens of sparse optimization. We propose a new solution method using variable splitting; in particular, we derive closed form updates in a highly efficient ADMM algorithm. Notably, no matrix products---let alone inversions---are required at any step. This reduces computational cost by orders of magnitude over existing methods, and the resulting algorithm is easily parallelizable and fairly insensitive to tuning parameters. A comparison to prior work is carried out in two applications to models of blood cell production and transposon evolution, showing that the proposed method is orders of magnitudes more scalable than existing work. 
\end{abstract}

\section{Introduction}
Continuous time Markov Chains (CTMCs) have been widely used to stochastically model population dynamics with diverse applications in the fields of genetics, epidemiology, finance, cell biology, and nuclear fission \citep{renshaw2015stochastic, allen2010introduction, paul2013stochastic}. Their popularity is owed to their flexibility, interpretability, and desirable mathematical properties. From the perspective of statistical inference, one of the most fundamental quantities characterizing a CTMC is its \textit{transition probabilities}, the conditional probabilities that a chain ends at a specific state given a starting state and finite time interval do not have closed form expressions. The set of all transition probabilities fully defines a process and forms the backbone of central quantities such as the likelihood function of discretely observed data from a CTMC \citep{guttorp2018stochastic}. Unfortunately, obtaining these transition probabilities is often computationally intensive, as it requires marginalization over an infinite set of endpoint-conditioned paths \citep{hobolth2009simulation}. 

Classically, this is achieved by computing the matrix exponential of the infinitesimal generator of the CTMC at the expense of having $O(N^3)$ runtime complexity, where $N$ is the size of the state space. Since this procedure is not suitable even for state spaces of moderate sizes, practitioners often rely on simplifying assumptions or sampling-based approaches via Markov Chain Monte Carlo (MCMC) algorithms \citep{rao2012fast, ross1987approximating, grassmann1977transient}, each with their own drawbacks. Many core frequentist and Bayesian inferential methods require an iterative evaluation of the observed data likelihood. 

In certain structured classes within CTMCs, alternative approaches are possible. Transition probabilities can be calculated explicitly for simple examples such as the Poisson process, and the class of linear birth-death processes provides another special case that admits closed-form solutions \citep{crawford2014estimation,KARLIN19751, lange2010applied}.

In this article, we focus on the case of multi-type branching processes, for which efficient numerical techniques have only recently been developed.
\citet{xu2015likelihood} make use of a method that expresses the probability generating function (PGF) as a Fourier series expansion. The generating functions are obtained as solutions to differential equation systems, and the transition probabilities can be recovered from fast numerical series inversion techniques, allowing maximum likelihood inference and Expectation-Maximization (EM) algorithms \citep{doss2013fitting}). 
However, these methods can also become costly for large systems: for instance, approximating the Fourier inversion formula using a Riemann sum requires ${O}(N^b)$ PGF evaluations, where $b$ is the number of types in the branching process and $N$ is the largest population size at the end of the desired transition probabilities, which entails millions of PGF evaluations even for a two-type process with populations in the thousands. As many likelihood-based procedures are iterative, scalable alternatives to this computation are critical to avoid a bottleneck that renders these methods out of reach. When sparsity is available in the probabilities, \citet{xu2015efficient} suggest the use of compressed sensing to reduce the number of computations at the expense of solving a sparse optimization problem using proximal gradient descent (PGD).   

To progress, we revisit the compressed sensing framework of \citet{xu2015efficient} from the lens of variable splitting methods. We show how the sparse optimization problem can be solved by orders of magnitude more efficiently via a careful implementation of the alternating direction method of multipliers (ADMM) algorithm \citep{everett1963generalized, eckstein1994some, boyd2011distributed}. 

Our approach operates on a vectorized version of the problem, avoiding large matrix inversions that prevent the previous PGD approach from applying to large-scale settings. Surprisingly, not only does variable splitting avoid inversions, but can be accomplished without calls even to matrix multiplication by clever use of the Fast Fourier Transform. Our method inherits desirable convergence and recovery guarantees and admits easily parallelizable implementations. Moreover, it is surprisingly inelastic to tuning parameters --- we find that successful performance is robust to a broad range of reasonable penalty parameters.  We validate these merits in detailed empirical examples, showcasing significant advantages over prior algorithms applied to a two-type branching model of hematopoiesis, the process of blood cell formation, and a model of transposon evolution from genetic epidemiology. 

\section{Markov Branching Processes}
A Markov branching process is a continuous-time Markov chain comprised of a collection of particles that proliferate independently, and whose reproduction or death follows a probability distribution. We consider continuous-time, multitype branching processes that take values over a discrete state space of non-negative integers where each particle type has its own mean lifetime and reproductive pattern \citep{lange2010applied, KARLIN19751}. For exposition, we focus on the two-type case, though the results generally apply.

Let $\pmb{X(t)}$ denote a linear, two-type branching process that takes values in a discrete state space $\Omega$, with $X_{i}(t)$ representing the number of particles of type $i$ present at time $t \geq 0$, where each particle of type $i$ at the end of its lifespan can produce $k$ particles of type 1 and $l$ particles of type 2 with instantaneous rates $a_{i}(k, l)$. The overall rates are multiplicative in the number of particles as a result of the linearity of the rate, which follows from the independence assumption. 

While they are defined by the instantaneous rates, the transition probabilities provide an alternate characterization of a branching process, also completely defining its dynamics:
\begin{equation}
    p_{\mathbf{x},\mathbf{y}}(t) = \mathbb{P}(\mathbf{X}(t+s)=\mathbf{y}|\mathbf{X}(s)=\mathbf{x}).
    \label{eq:1.1}
\end{equation}

These transition probabilities play a central role in statistical inference, either in the form of maximum likelihood estimation or within Bayesian methods, where likelihood calculations often enter in schemes such as Metropolis-Hastings to sample from the posterior density. Motivated by this, we now focus our attention on computing them in the class of continuous-time branching process.

\subsection{Generating function}
The probability generating function (PGF) for a two-type process is
\begin{align}
    \phi_{jk}(t,s_{1}, s_{2}; \pmb{\theta}) &:= \mathbb{E_{\pmb{\theta}}}(s_{1}^{X_{1}(t)}s_{2}^{X_{2}(t)}|X_{1}(0) = j, X_{2}(0) = k) \nonumber\\
    &= \sum_{l=0}^{\infty}\sum_{m=0}^{\infty}p_{(jk),(lm)}(t;\pmb{\theta})s_{1}^{l}s_2^{m} \label{eq:1.3},
\end{align}
with a natural extension existing for any $m$-type process. Given the instantaneous rates $a_{j}(k,l)$, the system of differential equations governing $\phi_{jk}$ can be derived using the Kolmogorov forward or backward equations as shown by \citet{bailey1991elements}. Once we have $\phi_{jk}$, the transition probabilities can be formally obtained by differentiating $\phi_{jk}$ in Equation \ref{eq:1.3} and normalizing by an appropriate constant,
\begin{equation}
  p_{(jk),(lm)}(t) = \frac{1}{l! m!}\frac{\partial^l}{\partial{s_{1}}}\frac{\partial^m}{\partial{s_{2}}} \phi_{jk}(t)\Bigr|_{\substack{s_{1}=s_{2}=0}}.
  \label{eq:1.4}
\end{equation}

However, in practice repeated numerical differentiation is a computationally intensive procedure and often becomes numerically unstable, especially for large $l,m$. To avoid this issue, we use a technique by \citet{lange1982calculation} that allows us to use the Fast Fourier transform (FFT) to calculate the transition probabilities $p_{(jk),(lm)}(t)$ that occur as coefficients of $s_{1}^{l}s_2^{m}$ in Equation \ref{eq:1.3}. We can map the domain $s_{1},s_{2} \in [0,1]\times[0,1]$ to the boundary of the complex unit circle by setting $s_{1} = e^{2\pi i\omega_{1}}, s_{2} = e^{2\pi i\omega_{2}}$ and view the PGF as a Fourier series
\begin{equation*}
    \phi_{jk}(t,e^{2\pi i\omega_{1}}, e^{2\pi i\omega_{2}}) = \sum_{l,m=0}^{\infty}p_{(jk),(lm)}(t)e^{2\pi il\omega_{1}}e^{2\pi im\omega_{2}}.
\end{equation*}
We can now compute the transition probabilities via the Fourier inversion formula, approximating the integral by a Riemann sum:
\begin{align}
    p_{(jk),(lm)}(t) &= \int_{0}^{1} \int_{0}^{1} \phi_{jk}(t,e^{2\pi i\omega_{1}}, e^{2\pi i\omega_{2}})\nonumber\\
    & \times e^{-2\pi il\omega_{1}}e^{-2\pi im\omega_{2}} d\omega_{1}d\omega_{2}\nonumber\\
    &\approx \frac{1}{N^{2}} \sum_{u=0}^{N-1}\sum_{v=0}^{N-1} \phi_{jk}(t,e^{2\pi iu/N}, e^{2\pi iv/N})\nonumber\\
    & \times e^{-2\pi ilu/N}e^{-2\pi imv/N}. \label{eq:1.5}
\end{align}

\section{Compressed Sensing Framework}
Compressed sensing (CS) is based on the principle that a sparse or compressible signal can be reconstructed, often perfectly, from a small number of its projections onto a certain subspace through solving a convex optimization problem. For our application, transition probabilities play the role of a target sparse signal of Fourier coefficients. CS enables us to restrict the necessary computations to a much smaller sample of PGF evaluations, which play the role of measurements used to recover the sparse signal.

In this setup, let $\mathbf{u} \in \mathbb{C}^{N}$ be an unknown sparse signal and let $\mathbf{\Psi} = [\psi_{1}, \psi_{2}, \ldots, \psi_{N}] \in \mathbb{C}^{N\times N}$ denote an orthonormal basis of $\mathbb{C}^{N}$. Then there exists a unique $\mathbf{s} \in \mathbb{C}^{N}$ such that
\begin{equation}
    \mathbf{u} = \sum_{i=1}^{N} \psi_{i}s_{i} = \mathbf{\Psi s}.
    \label{eq:2.1}
\end{equation}
If the number of non-zeros in $\mathbf{s}$, or $||s||_{0}$, is less than $K$, then $\mathbf{u}$ is said to be $K$-\textit{sparse} under $\mathbf{\Psi}$. We are interested in cases where $K<<N$ or $\mathbf{u}$ is highly compressible. \par 

Let $M \in \mathbb{Z}^{+}$ satisfy $K<M<<N$ and $\mathbf{u}$ be observed through a measurement
\begin{equation}
    \mathbf{b} = \mathbf{\Phi} \mathbf{u} = \mathbf{As}\in \mathbb{C}^{M},
    \label{eq:2.2}
\end{equation}
where $\mathbf{\Phi} \in \mathbb{C}^{M\times M}$ denotes a nonadaptive sensing matrix and $\mathbf{A = \Phi\Psi}$. Here nonadaptiveness means that $\mathbf{\Phi}$ does not depend on $\mathbf{u}$. As $M<<N$, this system is underdetermined and the space of solutions is an infinite affine space. However, in certain sparse settings,  \citet{candes2006compressive} proves that reconstruction can be accurately achieved by finding the most sparse solution among all solutions of Equation \ref{eq:2.2}, i.e.,
\begin{equation}
    \hat{\mathbf{s}} = \{\mathop{\arg \min}\limits_{\mathbf{s}} ||s||_{0} : \mathbf{As=b}\}.
    \label{eq:2.3}
\end{equation}
Due to the combinatorial intractability of the non-convex objective function, it is  impractical to solve this $\ell_{0}$ problem directly. Instead we use the $\ell_{1}$- relaxation as a proxy which results in a nice convex optimization problem, where we optimize the following unconstrained penalized objective
\begin{equation}
    \hat{\mathbf{s}} = \mathop{\arg \min}\limits_{\mathbf{s}} \frac{1}{2}||\mathbf{As} - \mathbf{b}||^{2}_{2} + \lambda ||\mathbf{s}||_{1},
    \label{eq:2.4}
\end{equation}
with $\lambda$ serving as the regularization parameter to enforce the sparsity of $\mathbf{s}$. \par 
Past works indicate that when $\mathbf{A}$ satisfies the \textit{Restricted  Isometry Property} (RIP) \citep{candes2005decoding}, \cite{candes2006robust}, the \textit{K}-sparse signal $\mathbf{u}$ or equivalently $\mathbf{s}$ can be reconstructed using only $M = CK \log N$ measurements for some constant $C$. The result assumes that the columns of $\boldsymbol{\phi}$ and $\boldsymbol{\psi}$ are not just incoherent pairwise, but \textit{K}-wise incoherent. The \textit{coherence} between $\boldsymbol{\phi}$ and $\boldsymbol{\psi}$ is defined as $\mu(\boldsymbol{\phi}, \boldsymbol{\psi}) = \sqrt{n}\mathop{ \max}\limits_{1 \leq i,j\leq n}|\left<\boldsymbol{\phi}_{i}, \boldsymbol{\psi}_{j}\right>|$. 

In practice, verifying RIP can be a demanding task;  \citet{candes2006robust} and \citet{donoho2006compressed} show that RIP holds with high probability in some Gaussian random matrix settings. In the focus of this paper, $\mathbf{A}$ is comprised of a random ``spike" basis playing the role of $\mathbf{\Phi}$ together with a Fourier domain representation $\mathbf{\Psi}$ \citep{rudelson2008sparse,candes2006robust}, which form a maximally incoherent pair \citep{candes2007sparsity}. We see how these components enter the derivation below, and later confirm the success this theory suggests via a thorough empirical study. 

\subsection{Higher dimensions}
We can easily extend this exposition focused on the vector valued case to higher-dimensional signals \citep{candes2006compressive}. To illustrate this, consider the $2D$ case where the sparse solution $\mathbf{S} \in \mathbb{C}^{N\times N}$ and the measurement
\begin{equation}
    \mathbf{B} = \mathbf{ASA}^{T} \in \mathbb{C}^{M\times M}
    \label{eq:2.5}
\end{equation}
are matrices instead of vectors. We may solve an equivalent problem
\begin{equation}
    \mathbf{\hat{S}} = \mathop{\arg \min}\limits_{\mathbf{S}} \frac{1}{2}||\mathbf{ASA}^{T} - \mathbf{B}||^{2}_{2} + \lambda ||\mathbf{S}||_{1}.
    \label{eq:2.6}
\end{equation}
Equivalently, this can also be represented in a vectorized framework with
\begin{equation*}
    \text{vec}(\mathbf{S}) = \Tilde{\mathbf{s}} \in  \mathbb{C}^{N^{2}}, \quad \text{vec}(\mathbf{B}) = \Tilde{\mathbf{b}} \in  \mathbb{C}^{M^{2}},
\end{equation*}
and we now pursue $\Tilde{\mathbf{b}} = \Tilde{\mathbf{A}}\Tilde{\mathbf{s}}$, with $\Tilde{\mathbf{A}} = \mathbf{A} \bigotimes \mathbf{A}$ being the Kronecker product of $\mathbf{A}$ with itself. It can be preferable to solve Equation \ref{eq:2.6} rather than the vectorized problem explicitly, as the number of entries in $\Tilde{\mathbf{A}}$ grows rapidly. However, we will show how working with vectorized forms together with clever implementations of the Fast Fourier Transform will provide the best of both worlds, avoiding matrix multiplication and inversion entirely.

\section{Transition Probabilities via ADMM}

Recall that we wish to compute the transition probabilities $p_{jk,lm}(t)$ for any $t>0$ and initial  $\mathbf{X}(0) = (j,k)$. Within the CS framework described in the previous section, $\mathbf{S} \in \mathbb{C}^{N\times N}$ is the matrix of transition probabilities with entries
\begin{equation*}
    \{\mathbf{S}\}_{l,m} = p_{jk,lm}(t).
\end{equation*}
Without the CS framework, one can directly obtain the transition probabilities using Equation \ref{eq:1.5} by first computing an equally sized matrix of PGF solutions

\begin{equation}
    \mathbf{\Tilde{B}} = \{\phi_{jk}(t,e^{2\pi iu/N}, e^{2\pi iv/N})\}_{u,v = 0}^{N-1} \in \mathbb{C}^{N \times N}.
\label{eq:3.1}
\end{equation}
Obtaining $\mathbf{\Tilde{B}}$ becomes computationally expensive for large $N$ values, especially when each PGF must be solved, for instance via numerically evaluating a differential equation. Given a way to compute $\mathbf{\Tilde{B}}$, we recover the transition probabilities by taking the Fast Fourier Transform (FFT). We can better understand how this fits within the CS framework with the help of matrix operations. We have $\mathbf{S}=\mathbf{F}\mathbf{\Tilde{B}}\mathbf{F^{T}}$, where $\mathbf{{F}} \in \mathbb{C}^{N\times N}$ denotes the Discrete Fourier Transform (DFT) matrix. Thus, the Inverse Discrete Fourier Transform (IDFT) matrix $\mathbf{{F}^{-1}}$ becomes the sparsifying basis $\mathbf{\Psi}$ mentioned in Equation \ref{eq:2.1}, and we have $\mathbf{\Tilde{B}}=\mathbf{\Psi S \Psi^{T}}$. 


If we expect $\mathbf{S}$ to have a sparse representation, we can employ a method to reconstruct $\mathbf{S}$ using a much smaller set of PGF evaluations arranged in the matrix $\mathbf{B} \in \mathbb{C}^{M\times M}$, corresponding to a subset of entries from $\mathbf{\Tilde{B}}$ selected uniformly at random. This much smaller matrix $\mathbf{\Tilde{B}}$ is a projection $\mathbf{B} = \mathbf{ASA}^{T} \in \mathbb{C}^{M\times M}$ in accordance with Equation \ref{eq:2.5}, where $\mathbf{A} \in \mathbb{C}^{M\times N}$ is obtained by selecting a subset of rows of $\mathbf{\Psi}$ that correspond to $\mathcal{J}$, the randomly sampled indices. This uniform sampling of rows is equivalent to multiplication by measurement matrix encoding the \textit{spike} (or standard) basis. Mathematically, we have $\mathbf{A = \Phi\Psi}$, with the rows of the measurement matrix $\mathbf{\Phi}_{j}(l) = \delta (j-l)$. Thus, uniformly sampling the indices $\mathcal{J}$ is optimal in our setting because the spike and Fourier bases are \textit{maximally incoherent} in any dimension \citep{candes2008introduction}. 

In the compressed sensing framework, we now require only computing this reduced matrix $\mathbf{B}$, which entails only a logarithmic proportion $|\mathbf{B}| \propto K \log|\mathbf{\Tilde{B}}|$ of PGF evaluations compared to the original problem. Thus, the problem of computing the transition probabilities $\mathbf{S}$ has been reduced to a signal recovery problem that can be solved by solving the optimization problem in Equation \ref{eq:2.6}. 

\subsection{Solving the convex problem using ADMM}

We use the Alternating Direction Method of Multipliers (ADMM) algorithm \citep{everett1963generalized, boyd2011distributed} to solve this problem efficiently. ADMM is useful for optimization problems where the objective function is of the form $\{f(\mathbf{x})+g(\mathbf{y}) : \mathbf{x}=\mathbf{y}\}$ with both $f(.)$ and $g(.)$ convex but not necessarily differentiable. 

We first introduce some notations before detailing the optimization routine. Recall that for our application, $\mathbf{U}$ is a two-dimensional unknown sparse signal being observed through a measurement matrix $\mathbf{B}$, and $\mathbf{F_{P}} = \mathbf{PF} \in \mathbb{C}^{M \times N}$ denotes the sensing matrix, where $\mathbf{P} \in \mathbb{R}^{M \times N}$ is a selection matrix containing $M$ rows of the identity matrix of order $N$. Similarly, we will denote $\mathbf{F^{-1}_{P}} = \mathbf{PF^{-1}} \in \mathbb{C}^{M \times N}$ where $\mathbf{{F}^{-1}}$ is the IDFT matrix.

To solve this problem efficiently using ADMM, we first split the optimization variable by introducing a new variable $\mathbf{Z}$ under an equality constraint with the variable of interest:
\begin{align}
        \min_{\mathbf{U},\mathbf{Z}} \quad & \frac{1}{2}||\mathbf{F_{P}^{-1}}\mathbf{U}\mathbf{(F_{P}^{-1}})^{H} - \mathbf{B}||^2_2 +\lambda||\mathbf{Z}||_1 \label{eq:3.2}\\
        \text{subject to} & \qquad \mathbf{U}=\mathbf{Z} .\nonumber
\end{align}

Next, we consider the Augmented Lagrangian
\begin{align}
        L_{\beta}(\mathbf{U}, \mathbf{Z}, \mathbf{Y}) &= \frac{1}{2}||\mathbf{F_P^{-1} U (F_P^{-1})}^{H} - \mathbf{B}||_{2}^2 +
        \lambda||\mathbf{Z}||_1 \nonumber\\ 
        &+  \mathbf{Y}^{T}(\mathbf{U}-\mathbf{Z}) + \frac{\beta}{2}||\mathbf{U}-\mathbf{Z}||_{2}^2,
        \label{eq:3.3}
\end{align}
where $\mathbf{Y}$ serves as the \textit{dual variable} enforcing $\mathbf{U} = \mathbf{Z}$. Now, the ADMM algorithm entails the following iteration:
\begin{equation}
\begin{split}
    \mathbf{U^{(k+1)}} &:= \argmin_\mathbf{U} L_{\beta}(\mathbf{U}, \mathbf{Z^{(k)}}, \mathbf{Y^{(k)}})\\
    \mathbf{Z^{(k+1)}} &:= \argmin_\mathbf{Z} L_{\beta}(\mathbf{U^{(k+1)}}, \mathbf{Z}, \mathbf{Y^{(k)}})\\
    \mathbf{Y^{(k+1)}} &:= \mathbf{Y^{(k)}} + \beta(\mathbf{U^{(k+1)}} - \mathbf{Z^{(k+1)}})\\
\end{split}
\end{equation}
Deriving the explicit update for the subproblem in $\mathbf{Z}$ here is straightforward, given by  the soft-thresholding operator
\begin{equation*}
    \mathbf{Z^{(k+1)}} = \text{SoftThresh}\Big(\mathbf{U^{(k+1)}} + \frac{\mathbf{Y^{(k)}}}{\beta}, \frac{\lambda}{\beta}\Big),
\end{equation*}
where SoftThresh$(\epsilon, \lambda/\beta)$ = $\max\{|\epsilon|- \lambda/\epsilon,0\}\cdot sign(\epsilon)$ for a constant $\epsilon \in \mathbb{R}^{+}$. To minimize $\mathbf{U}$, we set $\nabla_{U}\Big[L_{\beta}(\mathbf{U}, \mathbf{Z}, \mathbf{Y})\Big]$ equal to $0$,
\begin{align}
&\implies \nabla_{U}\Big[\frac{1}{2}||\mathbf{F_P^{-1} U (F_P^{-1})}^{H} - \mathbf{B}||_{2}^2\Big] + \nabla_{U}(\lambda||\mathbf{Z}||_1)\nonumber\\
&\qquad \qquad \qquad +\nabla_{U}(\mathbf{Y}^{T}(\mathbf{U}-\mathbf{Z})) +\nabla_{U}\Big[\frac{\beta}{2}||\mathbf{U}-\mathbf{Z}||_{2}^2\Big] = 0\nonumber,\\
&\implies \mathbf{(F_P^{-1})}^{H}\mathbf{(F_P^{-1}) U}\mathbf{(F_P^{-1})}^{H}\mathbf{(F_P^{-1})} - \mathbf{(F_P^{-1})}^{H}\mathbf{B}\mathbf{(F_P^{-1})}\nonumber\\
&\qquad \qquad \qquad +\mathbf{Y} + \beta(\mathbf{U - Z}) = 0\nonumber,\\
&\implies \mathbf{(F_P^{-1})}^{H}\mathbf{(F_P^{-1}) U}\mathbf{(F_P^{-1})}^{H}\mathbf{(F_P^{-1})} +  \beta\mathbf{U}\nonumber\\
&\qquad \qquad \qquad =\mathbf{(F_P^{-1})}^{H}\mathbf{B}\mathbf{(F_P^{-1})} + \beta\mathbf{Z} - \mathbf{Y}.\label{eq:3.4}
\end{align}
Notice that Equation \ref{eq:3.4} cannot be used to obtain a closed form update for $\mathbf{U}$. However, we can make progress by vectorizing, allowing us to rewrite Equation \ref{eq:3.4} as
\begin{align*}
(\mathbf{\tilde{F}_{P}^{-1}})^{H}\mathbf{\tilde{F}_{P}^{-1}} \mathbf{\tilde{u}^{(k+1)}} + \beta\mathbf{\tilde{u}^{(k+1)}} &= (\mathbf{\tilde{F}_{P}^{-1}})^{H}\mathbf{\tilde{b}} + \beta\Big(\mathbf{\tilde{z}^{(k)}} - \frac{\mathbf{\tilde{y}^{(k)}}}{\beta}\Big),
\end{align*}
where $\mathbf{\tilde{F}_{P}^{-1}} = \mathbf{(F_P^{-1})} \bigotimes \mathbf{(F_P^{-1})} \in \mathbb{C}^{M^{2} \times N^{2}}$ and $ \mathbf{\tilde{I}} = \mathbf{I} \bigotimes \mathbf{I} \in \mathbb{R}^{N^{2} \times N^{2}}$. This yields the update for $\mathbf{\tilde{u}}$
\begin{equation}
    \mathbf{\tilde{u}^{(k+1)}} = \Big[(\mathbf{\tilde{F}_{P}^{-1}})^{H}\mathbf{\tilde{F}_{P}^{-1}} + \beta\mathbf{\tilde{I}}\Big]^{-1} \Big[(\mathbf{\tilde{F}_{P}^{-1}})^{H}\mathbf{\tilde{b}} + \beta\Big(\mathbf{\tilde{z}^{(k)}} - \frac{\mathbf{\tilde{y}^{(k)}}}{\beta}\Big)\Big].
\label{eq:3.5}
\end{equation}

\textbf{Avoiding matrix operations:} Updating $\mathbf{\tilde{u}}$ na\"ively according to Equation \ref{eq:3.5} requires dense matrix multiplication and inversion, steps that would significantly add to the runtime of our algorithm in large-scale settings. 

To circumvent this, we may instead think of Equation \ref{eq:3.5} as the solution to the linear system
\begin{equation*}
    \mathbf{M}\mathbf{\tilde{u}^{(k+1)}} = \mathbf{a}, \quad \text{where}
\end{equation*}
$$\mathbf{M} = \Big((\mathbf{\tilde{F}_{P}^{-1}})^{H}\mathbf{\tilde{F}_{P}^{-1}} + \beta \mathbf{\tilde{I}} \Big) \in \mathbb{C}^{N^{2} \times N^{2}} \quad \text{and} $$
$$ \mathbf{a} = \Big((\mathbf{\tilde{F}_{P}^{-1}})^H \mathbf{\tilde{b}} + \beta\Big(\mathbf{\tilde{z}^{(k)}} - \frac{\mathbf{\tilde{y}^{(k)}}}{\beta}\Big)\Big) \in \mathbb{C}^{N^{2}}.$$ 
Multiplying $\mathbf{\tilde{F}^{-1}} = \mathbf{F^{-1}} \bigotimes \mathbf{F^{-1}} \in \mathbb{C}^{N^{2} \times N^{2}}$ to both sides of the above equation, we obtain
\begin{equation*}
    \mathbf{\hat{M}} \mathbf{\tilde{F}_{P}^{-1} \tilde{u}^{(k+1)}} = \mathbf{\hat{a}}, \quad \text{where}
\end{equation*}
\begin{equation}
\mathbf{\hat{M}} = \beta \mathbf{\tilde{I}} + \mathbf{\tilde{P}^{T}\tilde{P}}
   \quad\text{and}\quad 
\mathbf{\hat{a}} = \mathbf{\tilde{F}^{-1}}\beta \mathbf{\tilde{z}^{(k)}} + \mathbf{\tilde{P}^{T}\tilde{b}}.
\label{eq:3.8}
\end{equation}
The complete details of these derivations appear step-by-step in the Supplement. Note that $\mathbf{\hat{M}}$ is a diagonal matrix of order $N^{2}$ because both $\beta \mathbf{\tilde{I}}$ and $\mathbf{\tilde{P}^{T}\tilde{P}}$ are diagonal matrices of order $N^{2}$. Of course in practice we only compute and store the vector $\mathbf{\hat{m}}$ containing the $N^{2}$ diagonal elements of $\mathbf{\hat{M}}$. For exposition, let $N=4, M=2, \beta = 0.1$ and the uniformly sampled indices $\mathcal{J} = [1,2]$, then we can compute $\mathbf{\hat{m}}$ as,
\begin{align*}
    &\mathbf{\hat{m}_{1}} = 
          \beta\begin{bmatrix}
                    1 \\
                    \vdots \\
                    1
                \end{bmatrix}
          =
          \begin{bmatrix}
                    0.1 \\
                    \vdots \\
                    0.1
          \end{bmatrix}_{16 \times 1}
    \tilde{\mathbf{p}} =
          \begin{bmatrix}
           1 \\
           1 \\
           0 \\
           0
         \end{bmatrix}_{4 \times 1}
    \tilde{\mathbf{p}} \bigotimes \tilde{\mathbf{p}} = 
        \begin{bmatrix}
            1 \\
            1 \\
            0 \\
            0 \\
            1 \\
            1 \\
            (\mathbf{0})_{10}
          \end{bmatrix}_{16 \times 1}\\
\mathbf{\hat{m}} &= \mathbf{\hat{m}_{1}} + \tilde{\mathbf{p}} \bigotimes \tilde{\mathbf{p}}
    = \Big[1.1, 1.1, 0.1, 0.1, 1.1, 1.1, (\mathbf{0.1})_{10}\Big]^{T}_{1 \times 16}.\\
\end{align*}

On the other hand, since $\mathbf{\hat{a}} = \text{vec}(\mathbf{\hat{A}}) = \text{vec}(\mathbf{F^{-1}}\beta \mathbf{Z^{(k)}} + \mathbf{P^{T}BP})$, computing $\mathbf{\hat{a}}$ is equivalent to first computing $\mathbf{\hat{A}}$ and then vectorizing it. Thus, we can compute $\mathbf{\hat{m}}$, compute $\mathbf{\hat{a}}$ and then efficiently solve the linear system
\begin{equation}
    \mathbf{F}^{-1}\mathbf{U^{(k+1)}} = \text{vec}^{-1}(\mathbf{\hat{a}}\oslash\mathbf{\hat{m}}),
    \label{eq:3.9}
\end{equation}
where $\oslash$ is the \textit{Hadamard (elementwise) division} operator and $\text{vec}^{-1}$ is the inverse of the vectorization operator, i.e. reshaping $\mathbf{\hat{a}}\oslash\mathbf{\hat{m}}$ back into a $N \times N$ matrix. Thereafter, to isolate the optimization variable, we may recover $\mathbf{U^{(k+1)}}$ by ``cancelling out'' $\mathbf{F^{-1}}$ by applying the FFT on both sides. Doing so avoids explicitly multiplying matrices implied by \ref{eq:3.5}; details are mentioned in line $8$ of Algorithm \ref{alg:1}.

\textbf{Computational complexity:}
We now we analyze the per-iteration computational complexity of the subproblems in minimizing $\mathbf{Z}$ and $\mathbf{U}$. First, note the minimization of $\mathbf{Z}$ requires evaluating the soft-thresholding operator elementwise, and thus has a complexity of $O(k N^2)$, where \textit{k} is the number of ADMM iterations and \textit{N} is the maximum population size. 

Next, we consider the complexity of naively minimizing $\mathbf{U}$ if one were to use Equation \ref{eq:3.5} directly. Doing so would involve dense matrix multiplications to calculate $\mathbf{M}$ and $\mathbf{a}$, as well as matrix inversion to obtain $\mathbf{U}$. Calculating $\mathbf{M}$ has a complexity of $O(N^{4}M^{2})$ as it involves matrix multiplication of $\mathbf{\tilde{F}_{P}^{-1}} \in \mathbb{C}^{M^{2} \times N^{2}}$ with its complex conjugate. Calculating $\mathbf{a}$ has a complexity of $O(\max\{(N^{2}M^{2}, k N^{2}\})$ while inverting $\mathbf{M}$ has $O(N^{6})$ cost, thus incurring a total computational complexity of $O(\max\{N^{6}, \max\{N^{2}M^{2}, k N^{2}\}\}) = O(N^{6})$ for the $\mathbf{U}$ update.

Our discussion on reducing this complexity by avoiding matrix operations reduces this significantly, making use of $\mathbf{\hat{m}}$ and $\mathbf{\hat{A}}$ instead of $\mathbf{M}$ and $\mathbf{a}$ as described above. Forming $\mathbf{\hat{m}}$ has $O(N^{2})$ cost as it involves the calculation of $N^{2}$ non-zero diagonal elements of $\mathbf{\hat{M}}$. 
Then, computing $\mathbf{\hat{A}}$ requires the calculation of $\mathbf{P^{T}BP}$, $\beta \mathbf{Z^{(k)}}$ and its IDFT, which have complexities $O(M^{2})$, $O(k M)$, and $O(k N^{2} \log_{2} N)$ respectively. Therefore, the calculation of $\mathbf{\hat{A}}$ has a complexity of $O(\max\{M^{2}, k M, k N^{2} \log_{2} N\}) = O(k N^{2} \log_{2} N)$ which matches the cost of computing two-dimensional FFT required to obtain $\mathbf{U^{(k+1)}}$. Since this is the dominant complexity in solving the linear system described by Equation \ref{eq:3.9}, the overall complexity of the $\mathbf{U}$ update (and in turn the $\mathbf{Z}$ update) becomes $O(k N^{2} \log_{2} N)$, a dramatic reduction from $O(N^{6})$. 
 
It is worth mentioning that the total complexity of the algorithm can be further improved by leveraging the sparsity of the transition probability matrix and using sparse FFT (sFFT) techniques \citep{hassanieh2012simple}, which can reduce the complexity of the two-dimensional IDFT from $O(k N^{2} \log_{2} N)$ to $O(k M^{2} \log_{2} M)$. However, for our application, we use the Fastest Fourier Transform in the West (\texttt{FFTW}) \citep{frigo1998fftw} to implement FFT as the differences in execution times between implementing FFT using \texttt{FFTW} and sFFT are only appreciable for $N > 2^{17}$ \citep{hassanieh2012simple}, which is beyond the maximum population sizes considered in this paper.

\textbf{Execution time of ADMM:}

We further improve the wall-clock execution time of our algorithm by vectorizing intermediate steps (lines 2, 3 and 8 of Algorithm \ref{alg:1}) whenever possible, and making use of optimized Numpy libraries \citealp{harris2020array} and parallelized FFT in \texttt{FFTW} \citep{frigo1998fftw}. 

\subsection{ADMM Convergence and Stopping Criteria}

Having detailed the algorithm and having emphasized careful diagonalization tricks in the Fourier domain to significantly reduce runtime, we now discuss aspects related to convergence. We establish the convergence guarantees and stopping criteria of our ADMM algorithm. It is not difficult to show that our formulation inherits powerful standard convergence results. We begin by verifying assumptions behind one of the classical convergence theorems.

\begin{proposition}
The functions $f(\mathbf{U})$ and $g(\mathbf{Z})$ are closed, convex and proper. 
    \label{prop:1}
\end{proposition}

The functions $f(\mathbf{U})=\frac{1}{2}||\mathbf{F_{P}^{-1}}\mathbf{U}\mathbf{(F_{P}^{-1}})^{H} - \mathbf{B}||^2_2$ and $g(\mathbf{Z}) = \lambda||\mathbf{Z}||_1$ are closed, convex and proper by virtue of being the square of the $L_{2}$ norm and $L_{1}$ norm, respectively \citep{rudin1976principles, folland1999real}.

\begin{proposition}
\label{prop:2}
For $\beta=0$, $L_{0} = \lambda||\mathbf{Z}||_1 + \frac{1}{2}||\mathbf{F_{P}^{-1}U\mathbf{(F_{P}^{-1}})^{H}} - \mathbf{B}||_{2}^2 + \mathbf{Y}^{T}(\mathbf{U}-\mathbf{Z})$, the unaugmented Lagrangian has a saddle point for all $\mathbf{U},\mathbf{Z},\mathbf{Y}$. 
\end{proposition}

\begin{proposition}
    Under propositions \ref{prop:1} and \ref{prop:2}, the ADMM algorithm achieves:
    \begin{enumerate}
        \item Primal residual convergence: $\mathbf{R^{(k+1)}} = \mathbf{U^{(k+1)}} - \mathbf{Z^{(k+1)}} \to \mathbf{0}$  as $k \to \infty$.
        \item Dual residual convergence: $\mathbf{S^{(k+1)}} = \beta(\mathbf{Z^{(k+1)}} - \mathbf{Z^{(k)}}) \to \mathbf{0}$ as $k \to \infty.$
    \end{enumerate}
    \label{prop:3}
\end{proposition}

Proofs of Propositions \ref{prop:2} and \ref{prop:3} are given in the Supplement.

\subsubsection{Optimality Conditions and Stopping Criteria}
The necessary and sufficient optimality conditions for the optimization problem described in Equation \ref{eq:3.2} are defined by primal feasibility, 
\begin{equation}
    \mathbf{U^{*} + Z^{*} = 0},
    \label{eq:3.10}
\end{equation}
and dual feasibility,
\begin{align}
    0 &= \nabla\Big[\frac{1}{2}||\mathbf{F_P^{-1}U^{*}}\mathbf{(F_{P}^{-1}})^{H} - \mathbf{B}||_{2}^2\Big] + \mathbf{Y^{*}} \label{eq:3.11}\\
    0 &\in \partial{(\lambda||\mathbf{Z^{*}}||_1)}  - \mathbf{Y^{*}} \label{eq:3.12},
\end{align}
where $(\mathbf{U^{*},Z^{*},Y^{*}})$ is the saddle point of $L_{0}$, $\partial$ denotes the subdifferential operator \citep{rockafellar1970convex,hiriart2004fundamentals,borwein2006convex} due to $\lambda||\mathbf{Z^{*}}||_1$ being non-differentiable. Thus, Equations \ref{eq:3.10}-\ref{eq:3.12} constitute the three optimality conditions for the optimization problem described in Equation \ref{eq:3.2}. The last condition (\ref{eq:3.12}) always holds for $(\mathbf{U^{(k+1)}, Z^{(k+1)}, Y^{(k+1)}})$ whereas the other conditions give rise to the primal $\mathbf{R^{(k+1)}}$ and dual $\mathbf{S^{(k+1)}}$ residuals with proof in the Supplement; these residuals converge to zero as ADMM iterates according to Proposition \ref{prop:3}.

\begin{algorithm}[h]
\caption{Fast sparse reconstruction using ADMM}
\label{alg:1}  

    \SetKwInOut{Input}{Input}
    \SetKwInOut{Output}{Output}

\underline{function ADMM} $(\mathbf{B}, \beta, \lambda, \epsilon_{abs}, \epsilon_{rel}, N_{iter}, \mathbf{P}, \mathcal{J})$\;
    \Input{Measurements $\mathbf{B} \in \mathbb{C}^{M \times M}$ , list of uniformly sampled indices $\mathcal{J}$, stepsize $\beta$, regularization parameter $\lambda$, error thresholds $\epsilon_{abs}$ and $\epsilon_{rel}$, iteration max ${N_{iter}}$, matrix $\mathbf{P} \in \mathbb{R}^{M \times N}$.}
    \Output{A real-valued 2D matrix $\mathbf{\hat{S}}$} 
       
    \textbf{Initialize} $\mathbf{\hat{m}_{1}}$ with $\mathbf{1_{N^{2} \times 1}}$, $\mathbf{\tilde{p}}$ with $\mathbf{0_{N \times 1}}; \mathbf{\tilde{p}}[\mathcal{J}] = 1$;\\
    set $\mathbf{\hat{m}} = \mathbf{\hat{m}_{1}} + \tilde{\mathbf{p}} \bigotimes \tilde{\mathbf{p}}$;\\
    \textbf{Initialize U, Z} $\mathbf{\hat{A}}$ with $\mathbf{0_{N \times N}}$;\\ 
    $\mathbf{\hat{A}}[\mathcal{J},\mathcal{J}] = \mathbf{B}$;\\
    \While{k < $N_{iter}$}
      {
      $\mathbf{\hat{a}} = \text{vec}(\mathbf{\hat{A}} + IFFT_{2D}(\beta \mathbf{Z}^{(k)}))$\;

      $\mathbf{U^{(k+1)}} = FFT_{2D}[\text{vec}^{-1}(\mathbf{\hat{a}}\oslash\mathbf{\hat{m}})]$\;

      $\mathbf{Z^{(k+1)}} =$ SoftThresh$(\mathbf{U^{(k+1)}} + \mathbf{Y^{(k)}}/\beta, \lambda/\beta)$\;

        $\mathbf{Y^{(k+1)}} = \mathbf{Y^{(k)}} + \beta(\mathbf{U^{(k+1)}} - \mathbf{Z^{(k+1)}})$\;
        
        \tcp{Stopping criteria using primal and dual tolerances}
        $\epsilon_{pri} = N\epsilon_{abs} + \epsilon_{rel} \max(||\mathbf{U^{(k)}}||_{2}, ||\mathbf{Z^{(k)}}||_{2})$\;
        $\epsilon_{dual} = N\epsilon_{abs} + \epsilon_{rel} ||\mathbf{Y^{(k)}}||_{2}$\;
        \If{$||\mathbf{R^{(k)}}||_{2} < \epsilon_{pri}$ \text{and} $||\mathbf{S^{(k)}}||_{2} < \epsilon_{dual}$}{
            return $\mathbf{\hat{S}} = \mathbf{U^{(k)}}/N^2$\;
        }
      }
      return $\mathbf{\hat{S}} = \mathbf{U^{(k)}}/N^{2}$\;
\end{algorithm}

We can thus use these residuals to put a bound on the objective suboptimality of the current point, $\frac{1}{2}||\mathbf{F_{P}^{-1}U^{(k)}}\mathbf{(F_{P}^{-1}})^{H} - \mathbf{B}||_{2}^2+\lambda||\mathbf{Z^{(k)}}||_1 - q^{*}$, where $q^{*}$ is the optimal value that the objective function given in Equation \ref{eq:3.2} will converge to for $k \to \infty$. The following equation appearing in the proof of convergence in the Supplement,
\begin{align*}
    \frac{1}{2}||\mathbf{\tilde{F}_{P}^{-1}\tilde{u}^{(k)}} & -  \mathbf{\tilde{b}}||_{2}^2+\lambda||\mathbf{\tilde{z}^{(k)}}||_1 - q^{*}\\
    \leq & -(\mathbf{\tilde{y}^{(k)}})^{T}\mathbf{\tilde{r}^{(k)}} + (\mathbf{\tilde{u}^{(k)}} - \mathbf{\tilde{u}^{*}})^{T}\mathbf{\tilde{s}^{(k)}},
\end{align*}
suggests that when the primal and dual residuals are small, the objective suboptimality must also be small, thereby motivating the use of the following stopping criterion:
\begin{equation*}
||\mathbf{R^{(k)}}||_{2} \leq \epsilon_{pri}
   \quad\text{and}\quad 
||\mathbf{S^{(k)}}||_{2} \leq \epsilon_{dual}.
\end{equation*}

Here $\epsilon_{pri} > 0$ and $\epsilon_{dual} > 0$ are the tolerances for the primal (\ref{eq:3.10}) and dual feasibility (\ref{eq:3.11}) conditions respectively. We choose the following tolerances for robustness analysis,
\begin{align}
    \epsilon_{pri} &= N\epsilon_{abs} + \epsilon_{rel} \max(||\mathbf{U^{(k)}}||_{2}, ||\mathbf{Z^{(k)}}||_{2}), \label{eq:3.13}\\
    \epsilon_{dual} &= N\epsilon_{abs} + \epsilon_{rel}, ||\mathbf{Y^{(k)}}||_{2}, \label{eq:3.14}
\end{align}
where N is the dimension of $\mathbf{U}$, $\epsilon_{abs}$ and 
$\epsilon_{rel}$ are the absolute and relative tolerances respectively, typically chosen between $10^{-2}$ to $10^{-4}$ depending on the application.

\section{Application and Empirical Study}
We propose to implement the framework described in the previous section to recover the transition probabilities for a two-type branching process; the idea and the framework can be easily extended to multi-type branching processes as well. As an outline, we calculate the full transition probability matrix $\mathbf{S}$ for a two-type branching process with known rates, randomly select a subset of the PGF evaluations $\mathbf{B}$, and use ADMM along with CS to recover the transition probabilities $\mathbf{\hat{S}}$ while also exploring the robustness of the ADMM algorithm with respect to the stepsize $\beta$, and the regularization parameter $\lambda$. Once satisfied with the robustness, we then proceed to compare the runtimes of recovering the transition probabilities using ADMM algorithm with the runtimes of PGD algorithm for similar errors.

We describe the two model applications used to generate data in our study.

\subsection{Two-compartment hematopoiesis model}
Hematopoiesis is a process of continuous formation and turnover of blood cells in order to fulfill the everyday demands of the body. At the origin of hematopoiesis is hematopoietic stem cell (HSC) which has two fundamental features: the first being the ability to self-renew, a divisional event which results in the formation of two HSCs from one HSC and the second being the ability for multipotent differentiation into all mature blood lineages. 
\par
We can stochastically model this biological phenomenon as a two-type branching process \citep{catlin2001statistical}. Under such representation, $X_{1}$ and $X_{2}$, the type one and type two particle populations correspond to HSCs and progenitor cells, respectively. With the same parameters as denoted in Figure \ref{fig:S2} of the Supplement, the non-zero instantaneous rates that define the process are,
\begin{align}
    a_{1}(2,0) &= \rho,  a_{1}(0,1) = \nu,  a_{1}(1,0) = -(\rho + \nu), \notag\\
    a_{2}(0,0) &= \mu, a_{2}(0,1) = -\mu.
    \label{eq:4.1}
\end{align}

Now that we have the instantaneous rates, we can derive the solutions for its PGF, defined in Equation \ref{eq:1.3}, and subsequently obtain the transition probabilities using Equation \ref{eq:1.5} with the details included in the Supplement. Note that the cell population in this application can reach the order of tens of thousands, thereby motivating the splitting methods approach to compute transition probabilities in an efficient manner.

\subsection{Birth-death-shift model for transposons}
For our second application, we look at the birth-death-shift (BDS) process proposed by \cite{rosenberg2003estimating} to model evolutionary dynamics of the genomic mobile sequence elements known as \textit{transposons}. Each transposon can either produce a new copy that can move to a new genomic location, shift to a different genomic location, or be completely removed from the genome, independently of all other transposons with per-particle instantaneous rates $\gamma, \sigma, \delta$ and overall rates proportional to the total number of transposons. The $IS6110$ transposon in the \textit{Mycobacterium tuberculosis} genome from a San Francisco community study dataset \citep{cattamanchi200613} was used to estimate these evolutionary rates by \cite{rosenberg2003estimating}. A two-type branching process can be used to model the BDS process over any finite observation interval \citep{xu2015likelihood}, where $X_{1}$ and $X_{2}$ represent the numbers of initially occupied genomic locations and newly occupied genomic locations, respectively, thus capturing the full dynamics of the BDS process.  

The two-type branching process representation of the BDS model has the following non-zero rates
\begin{align}
    a_{1}(1,1) &= \gamma,  a_{1}(0,1) = \sigma,  a_{1}(0,0) = \delta, \notag\\
    a_{1}(1,0) &= -(\gamma + \sigma + \delta), a_{2}(0,2) = \gamma, \notag\\
    a_{2}(0,1) &= -(\gamma + \delta), a_{2}(0,0) = \delta. 
    \label{eq:4.3}
\end{align}

\begin{proposition}
    The PGF of the BDS model described in Equation \ref{eq:4.3} is given by $\phi_{j,k} = \phi_{1,0}^{j} \phi_{0,1}^{k}$ by particle independence where,
    \begin{align}
        \phi_{0,1} (t, s_{1}, s_{2}) &= 1 + \notag\\
        &\Big[\frac{\gamma}{\delta - \gamma} + \Big(\frac{1}{s_{2}-1} + \frac{\gamma}{\gamma - \delta}\Big)e^{(\delta - \gamma)t}\Big]^{-1}, \notag\\
        \frac{d}{dt} \phi_{1,0} (t, s_{1}, s_{2}) &= \gamma \phi_{1,0} (t, s_{1}, s_{2}) \notag\\
        &+ \sigma  \phi_{0,1} (t, s_{1}, s_{2}) + \delta - (\gamma + \sigma + \delta)s_{1}).
        \label{eq:4.4}
    \end{align}
\end{proposition}

For our application, we calculate the transition probabilities $\mathbf{S}$ for maximum population sizes $N=2^{6}, 2^{7}, \ldots, 2^{10}$, given the time intervals $t$, branching process' rate parameters $\pmb{\theta}$ and initial population size $\mathbf{X(0)}$. Based on biologically sensible rates and observation times scales of data from previous hematopoiesis studies \citep{catlin2001statistical,golinelli2006bayesian,fong2009bayesian}, we set per-week branching rates and the observation time as $\boldsymbol\theta_{HSC} = [\rho, \nu, \mu] = [0.125, 0.104, 0.147]$ and $t=1$ week respectively. Similarly for the BDS application, based on previously estimated rates in \cite{xu2015likelihood} and the average length between observations in the tuberculosis dataset from San Francisco \citep{cattamanchi200613}, we set the per year event rates and the observation time as $\boldsymbol\theta_{BDS} = [\gamma, \sigma, \delta] = [0.016, 0.004, 0.019]$ and $t=0.35$ years respectively. In each case, we computed $M^{2} = (\min(\lfloor \sqrt{\log N K10} \rfloor, \lfloor \sqrt{N} - \sqrt{N}/5\rfloor)^{2}$ total random measurements to obtain $\mathbf{B}$, which will be used to recover the transition probabilities. 

\begin{figure*}[t]
\begin{subfigure}{0.33\textwidth}
  \includegraphics[width=.99\linewidth]{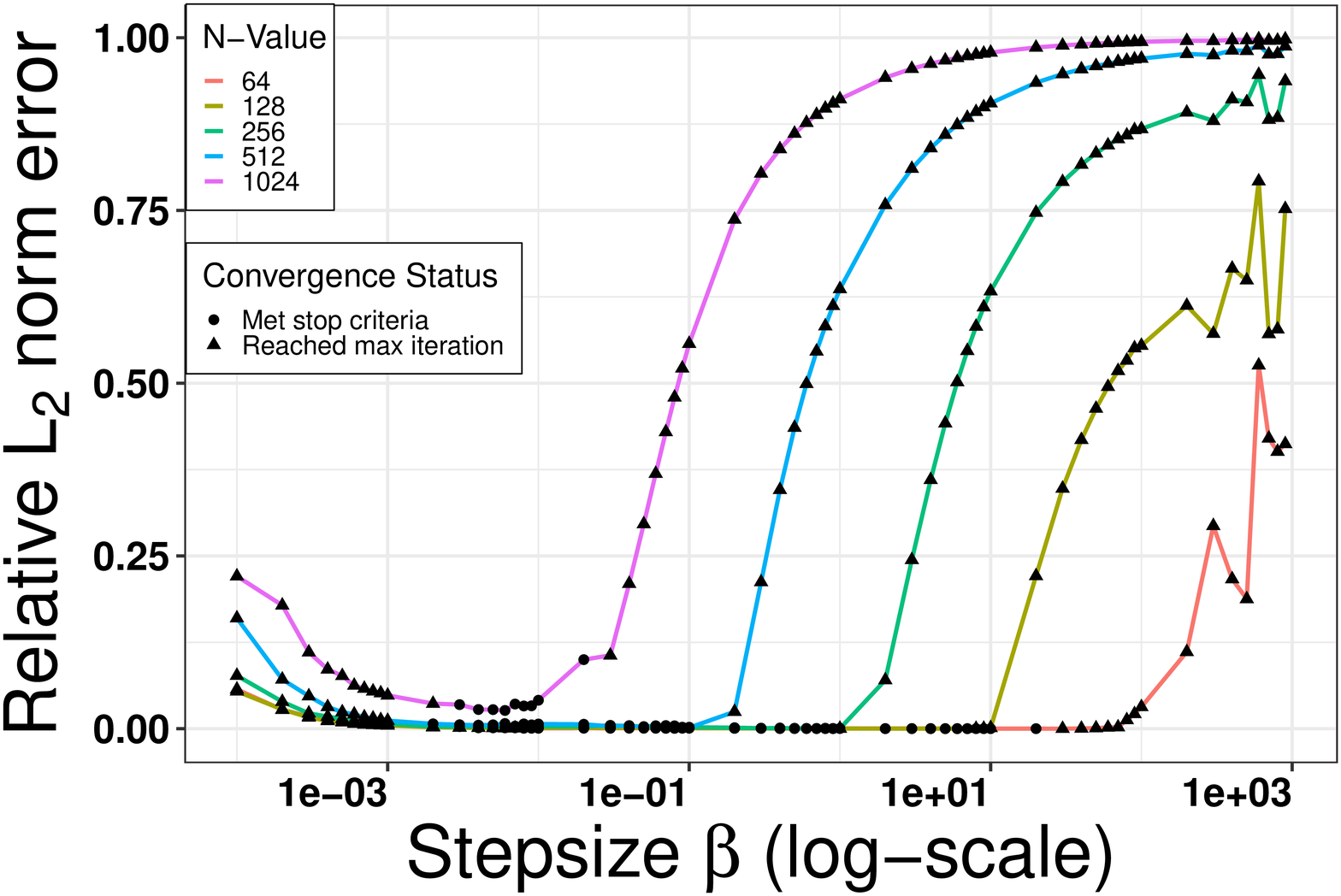}  
  \caption{}
  \label{fig:1a}
\end{subfigure}
\begin{subfigure}{0.33\textwidth}
  \centering
  \includegraphics[width=.99\linewidth]{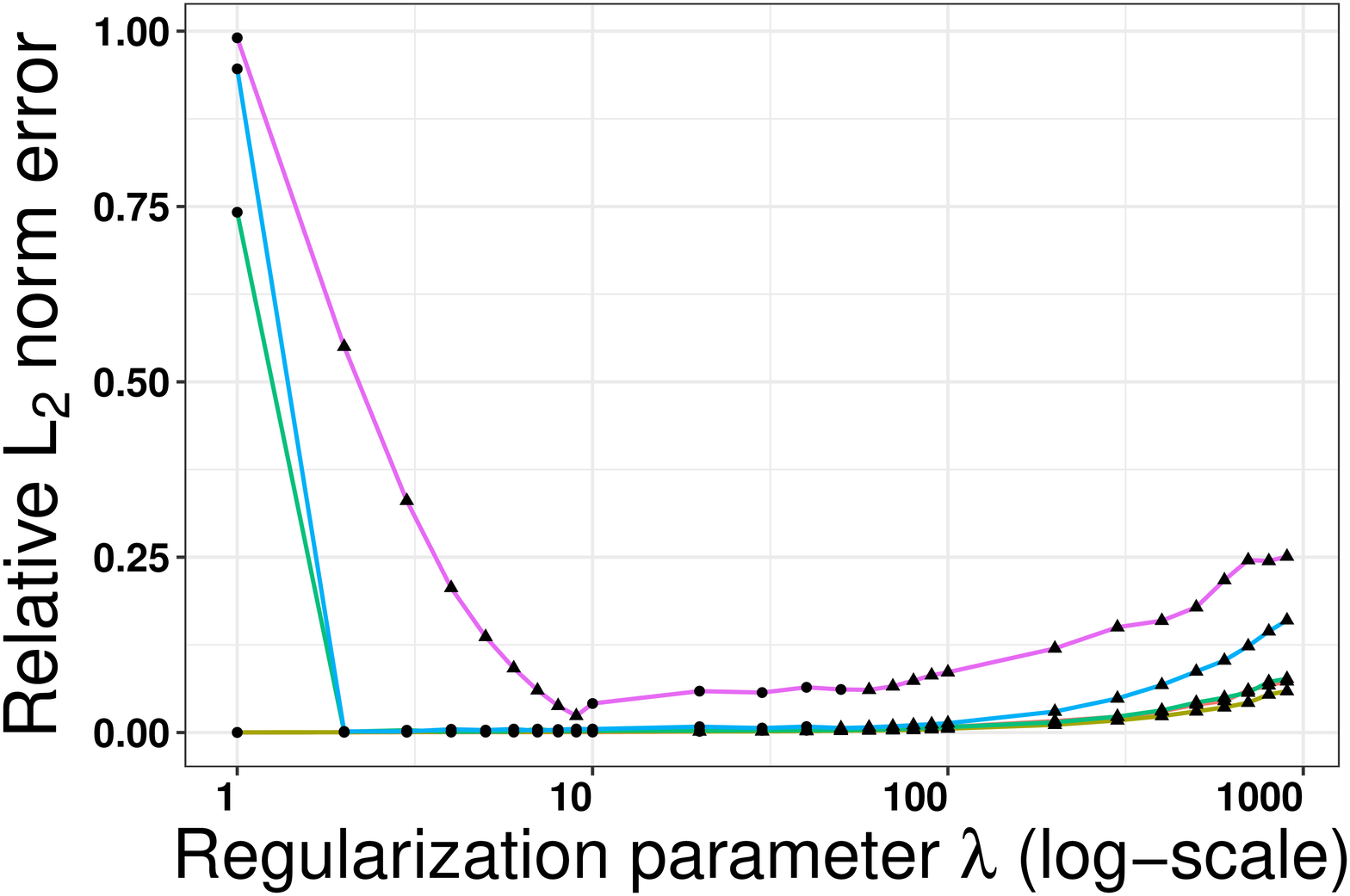}  
  \caption{}
  \label{fig:1b}
\end{subfigure}
\begin{subfigure}{0.36\textwidth} \vspace{-10pt}
  \centering
  \includegraphics[width=0.85\linewidth]{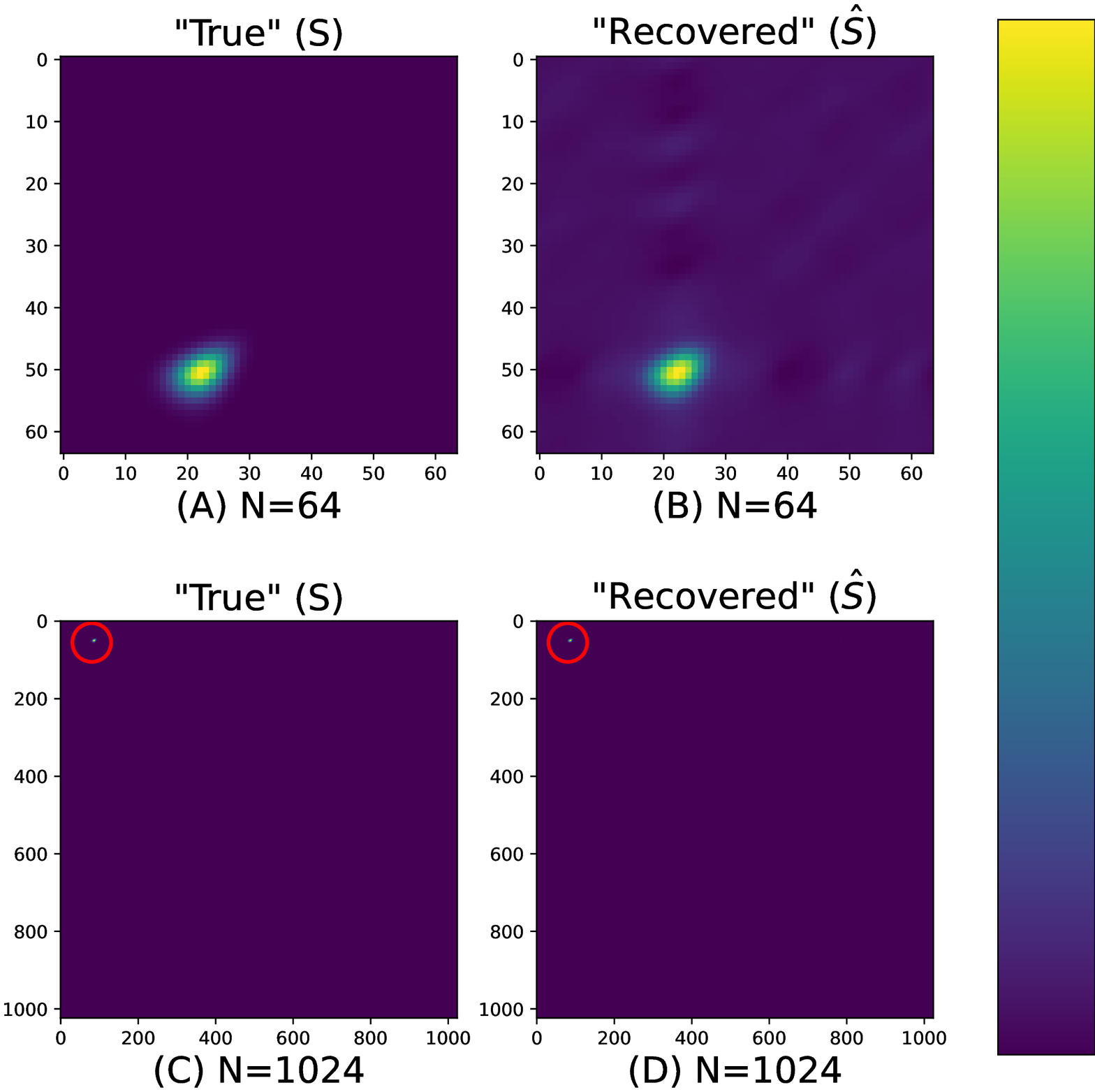}  
  \caption{}
  \label{fig:1c}
\end{subfigure}
\caption{Relative $L_{2}$ norm error, $\epsilon^{L_{2}}_{rel}$ for (a) varying stepsizes $\beta$ and (b) varying regularization parameter $\lambda$ under varying settings of $N$, HSC model. The successful convergence of ADMM algorithm is observed over a large range of stepsizes and $\lambda$. c) The sparse ``true'' transition probability matrix (\textbf{S}) along with the recovered transition probability matrix $(\pmb{\hat{S}})$ for $N=64$ and $N=1024$ for the HSC model using ADMM, with red circles highlighting non-zero transition probabilities.} \vspace{-5pt}
\label{fig:1}
\end{figure*}

\subsection{Robustness of ADMM algorithm}
We begin by investigating the robustness of our ADMM algorithm under a broad range of parameter settings. We consider varying the stepsize $\beta$ and the regularization parameter $\lambda$ for each $N$, and repeat this procedure across all $N$ values for the HSC model while examining the effects of these variations on the relative $L_{2}$ errors $\epsilon^{L_{2}}_{rel}$ defined 
\begin{equation*}
    \epsilon^{L_{2}}_{rel} = ||\mathbf{\hat{S}} - \mathbf{S}||_{2}/ ||\mathbf{S}||_{2},
\end{equation*}
where $||.||$ indicates the $L_{2}$ norm, $\pmb{S}$ and $\pmb{\hat{S}}$ represent the ``true'' and recovered transition probabilities respectively. 

In more detail, we start by fixing $\lambda=0.5(\log M)$ and vary $\beta$ in equal intervals from $10^{-4}$ to $9\times 10^{2}$ while recording and subsequently plotting $\epsilon^{L_{2}}_{rel}$ corresponding to each $\beta$. Thereafter, we fix $\beta = 0.08$, vary $\lambda$ from $1$ to $10^{3}$ and plot $\epsilon^{L_{2}}_{rel}$ corresponding to each $\lambda$. The primal $\epsilon_{pri}$ and dual $\epsilon_{dual}$ feasibility tolerances are given by Equations
\ref{eq:3.13} and \ref{eq:3.14} respectively with $\epsilon_{abs} = \epsilon_{rel} = 10^{-3}$.

Figure \ref{fig:1} displays these results. The algorithm is fairly inelastic to both the tuning of the stepsize $\beta$ as well as the penalty parameter $\lambda$: it achieves low errors conveying successful signal recovery for a wide range of values, which is visually evident even on the log scale. This is promising as our method may enter as a subroutine in optimization frameworks when the scale of these parameters is unknown, and must be learned throughout an iterative scheme. 

\subsection{Comparing the performance of PGD and ADMM}
Having confirmed that the method is not too sensitive to tuning, we compare its performance to the PGD algorithm proposed in \citet{xu2015efficient}. In order to compare the performance we first compute sets of transition probabilities $\textbf{S}$ of both HSC and BDS models using the full set of PGF solution measurements $\mathbf{\Tilde{B}}$ as described in Equation \ref{eq:3.1}.

\begin{figure}[H]
 \centering
  \includegraphics[width=0.45\linewidth]{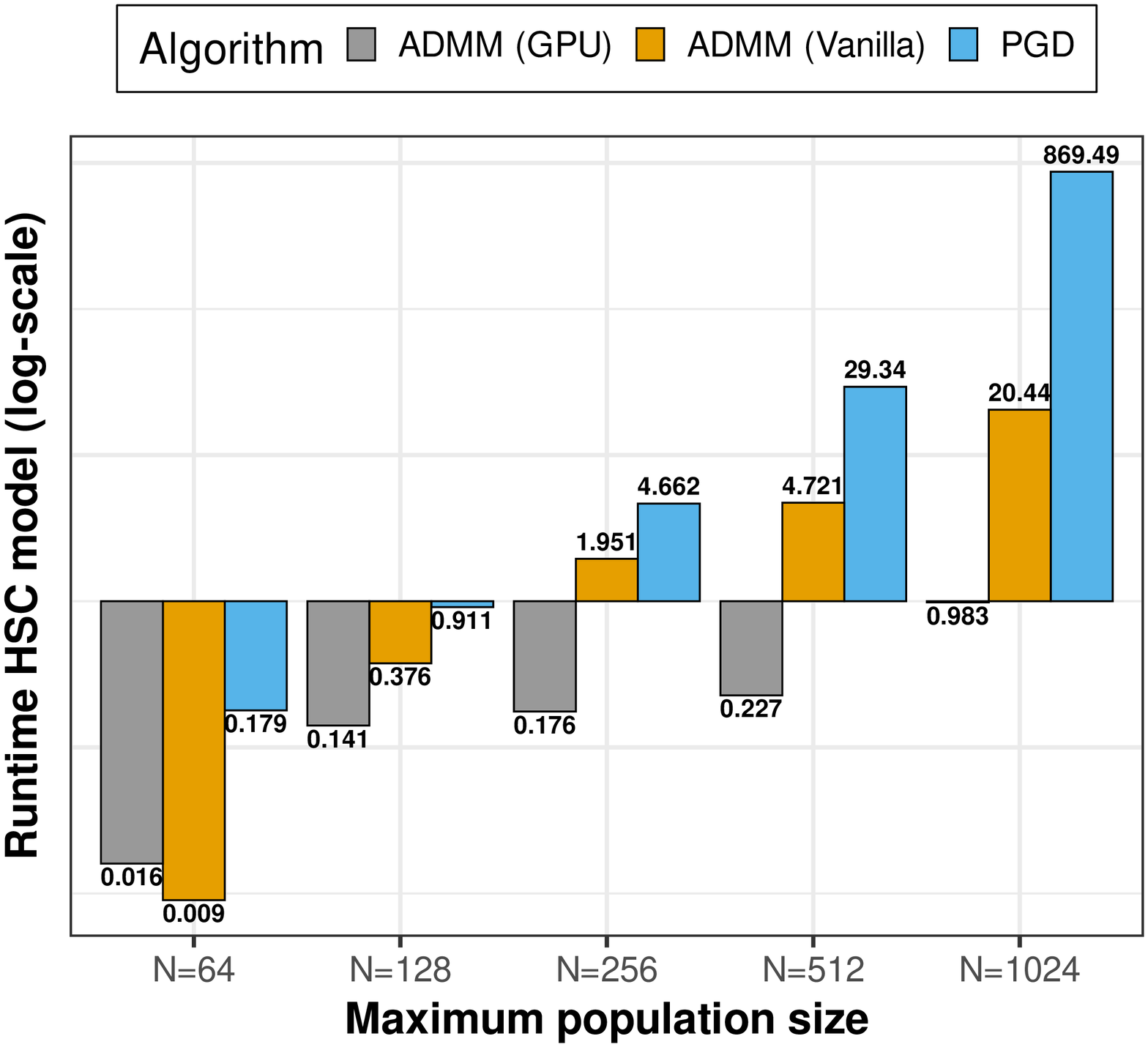}  
  \includegraphics[width=0.45\linewidth]{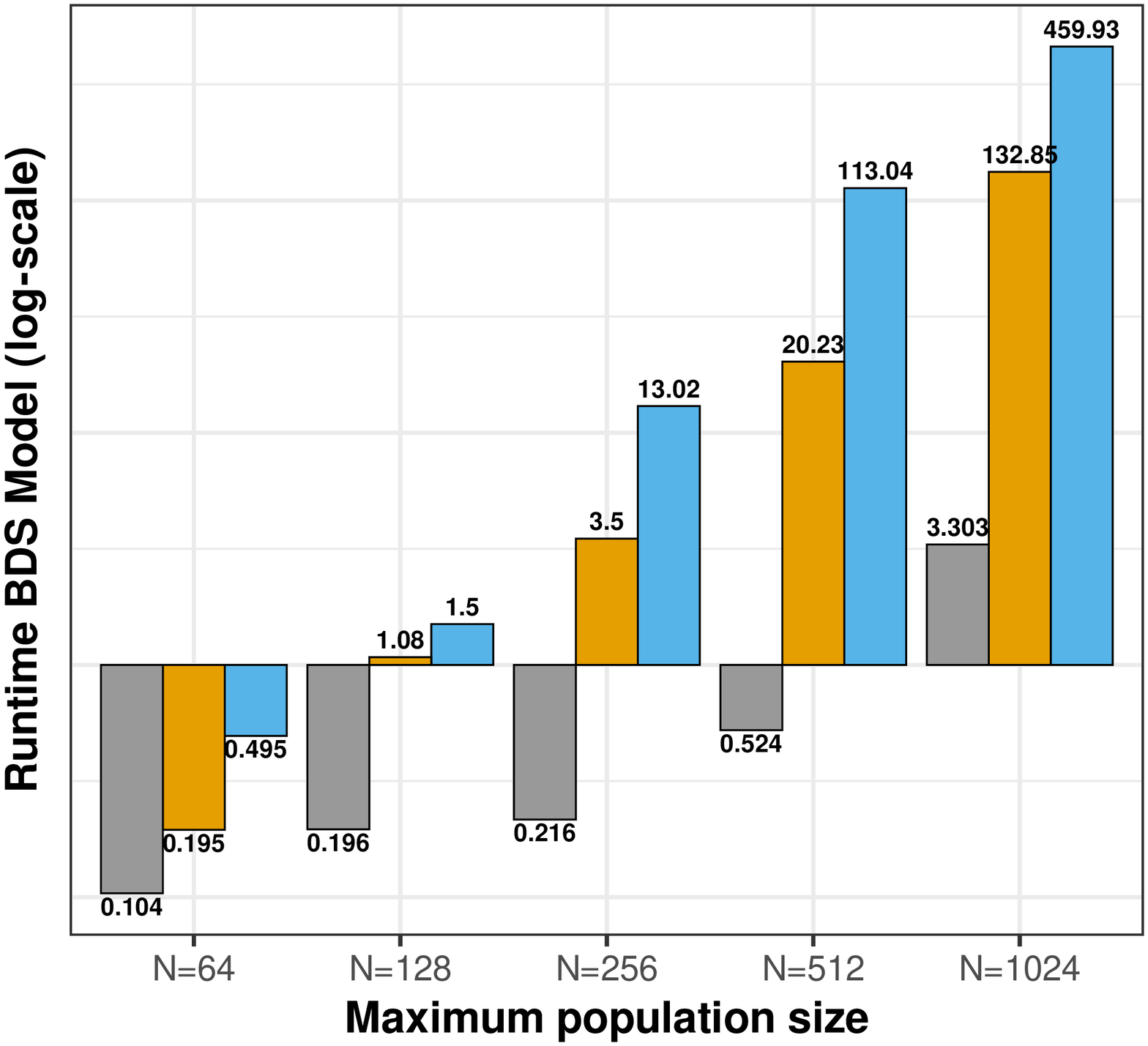}  
\caption{A comparison of median running times for different algorithms and maximum population sizes for the HSC model and  BDS model respectively. For all population sizes $N$ considered under each model, ADMM algorithm outpaces PGD algorithm under comparable or lower errors at convergence. Note results are plotted on the log-scale; this discrepancy becomes more pronounced as the size of the problem grows.}
\label{fig:3}
\end{figure}

Following \citet{xu2015efficient} we set the regularization parameters as $\lambda^{PGD}_{HSC} = \sqrt{\log M}$ and $\lambda^{PGD}_{BDS} = \log M$ for the PGD algorithm while keeping $\lambda = 0.5(\log M)$ for the ADMM algorithm for both the applications. It is promising that this simple heuristic for choosing the parameters leads to successful performance across \textit{all settings we consider}, further validating our empirical validation of robustness to tuning. Figure \ref{fig:1c} illustrates the sparse solution and provides a visual demonstration of the accuracy of reconstructed solutions using our ADMM approach, with the estimated matrix $\hat{\mathbf{S}}$ essentially identical to the ground truth.

For each value of N, we repeat this procedure five times under randomly sampled indices $\mathcal{J}$, and recover the transition probabilities $\mathbf{\hat{S}}$ using only a subset of PGF solution measurements $\mathbf{B}$ for both PGD and ADMM algorithms. We report median runtimes over the five trials in a fair conservative comparison, in that we ensure the measure of accuracy $\epsilon^{L_{2}}_{rel}$ under our proposed method is at least as good as PGD. To match errors across the algorithms, we choose 
$$\epsilon_{pri} = D_{1} \epsilon_{abs} + \epsilon_{rel} \max(||\mathbf{U^{(k)}}||_{2}, ||\mathbf{Z^{(k)}}||_{2}),$$
$$\epsilon_{dual} = D_{2} \epsilon_{abs} + \epsilon_{rel}, ||\mathbf{Y^{(k)}}||_{2}, $$
where $D_{1}, D_{2}$ are $\{N^{2}, N^{5}\}$ and $\{N^{2}, N^{2}\}$ for the HSC and BDS models respectively, complete details summarized in Table S1 of the Supplement. Since our ADMM algorithm largely depends on the FFT, a GPU based algorithm was also implemented to show that the algorithm can be accelerated drastically through straightforward parallelization, thus making it highly scalable for large data; further details on the GPU version are included in the Supplement.

The median runtimes of PGD, ADMM (Vanilla), and ADMM (GPU) for both HSC and BDS models are reported in Figure \ref{fig:3} while the median relative $L_{2}$ norm errors, $\epsilon^{L_{2}}_{rel}$ are reported in the Supplement. As evident from Figure  \ref{fig:3}, the ADMM algorithm consistently outperforms the PGD algorithm across all $N$ values for both the HSC and BDS models. For instance, with $N=1024$, there is a $97.65\%$ and $71.12\%$ reduction in the runtimes of ADMM (Vanilla) algorithm for the HSC model and BDS models respectively while the GPU implementation of ADMM is about $21$ times faster than ADMM (vanilla) and about $885$ times faster than PGD algorithm for the HSC model. Similarly, for $N=1024$ the ADMM (GPU) algorithm is about $40$ times faster than ADMM (vanilla) and about $140$ times faster than PGD algorithm for the BDS model.

\paragraph{Computing Infrastructure}
The experiment is conducted on Google Colaboratory\footnote{https://colab.research.google.com/} platform that claimed to have a RAM of 12.6 GB, a single-core Xeon CPU of 2.20 Ghz (No Turbo Boost) and a Tesla K80 with 2496 CUDA cores and 12GB GPU memory as well. The GPU version of FFT was implemented in CuPy \citep{cupy_learningsys2017}, a Numpy-like API library for CUDA with matched parameters in both the vanilla and GPU implementations of ADMM. The code and data used to implement the ADMM algorithm can be accessed from \url{https://github.com/awasthi-12/Transition_Prob_via_ADMM/}.

\section{Discussion}
We revisit the computational challenge of computing transition probabilities of large-scale multi-type branching processes. Focusing on the two-type setting, we derive a novel application of ADMM, showing how variable splitting can significantly accelerate methods to compute transition probabilities within a compressed sensing paradigm. Through a suite of experiments, we validate not only the robustness of ADMM with respect to the stepsize $\beta$ and the regularization parameter $\lambda$, but also its superior performance over previous attempts utilizing PGD. The advantages of the proposed method become especially pronounced as the population size $N$ increases. We also show through thoughtful algorithmic considerations that the primal variable ``$\mathbf{U}$''-update in ADMM can be carried out without matrix inversions. In particular, mathematically manipulating the expressions to leverage the FFT, the dominant complexity requires $O(N^{2} \log_{2}N)$ flops, dramatically increasing the scale of problems the method can consider. Lastly, a GPU-based parallel implementation further reduces the runtimes over the standard version.

In many realistic data settings where these stochastic population models apply, it is natural to expect sparsity in the support of transition probabilities. We have been able to achieve precise results in two such relevant examples under realistic parameter settings from scientific literature. Importantly, not only can the relevant quantities be computed efficiently, but in a way that succeeds over a wide range of tuning parameters. This robustness is crucial when embedding these methods as subroutines within inferential schemes such as maximum likelihood estimation \citep{doss2013fitting,xu2015likelihood}. Recall that for a Markov process $\mathbf{X}(t)$ observed at times $\{t_{1}, \ldots t_{T}\}$, the likelihood function of observed data is a product of its transitions
\begin{equation*}
    \mathcal{L(\mathbf{X};\boldsymbol\theta}) = \prod_{i=1}^{T-1}p_{\mathbf{X}(t_{i}),\mathbf{X}(t_{i+1})}(t_{i+1} - t_{i}; \boldsymbol\theta).
\end{equation*}
Optimizing this likelihood numerically often entails iterative schemes, and similar bottlenecks arise under other estimators as well as Bayesian approaches where the likelihood appears in routines such as Metropolis-Hastings ratios \citep{guttorp2018stochastic,stutz2022computational}. The robustness, efficiency, and accuracy of the proposed method make it well-suited to open the door to likelihood-based inference in previously intractable settings.

\section*{Acknowledgments}
This work was partially supported by NSF grants DMS-2230074 and PIPP-2200047. We thank Galen Reeves for helpful discussions and Yiwen Wang for early contributions to the code.

\bibliographystyle{abbrvnat}
\bibliography{references}

\begin{appendices}
\renewcommand{\theequation}{A.\arabic{equation}}
\setcounter{equation}{0}

\renewcommand{\thefigure}{S.\arabic{figure}}
\setcounter{figure}{0}
\renewcommand{\thetable}{S.\arabic{table}}
\setcounter{table}{0}

\section{Derivations and Proofs}
\subsection{Discrete Fourier transform matrix}
The $N\times N$ discrete Fourier transform (DFT) matrix $\mathbf{F}$ has entries $\{\mathbf{F}\}_{j,k} = \frac{1}{\sqrt{N}}(\omega)^{jk}$, where $\omega = e^{2\pi i/N}$ and $j,k = 0,1,\dots N-1$. As mentioned in the main text, recall that the Hermitian is given by its conjugate transpose: $\mathbf{F}^{H} = \big(\pmb{\mathbf{\overline{F}}}\big)^{T}$. Now let $\mathbf{x}$ be a vector with $\pmb{\hat{x}} = \mathbf{F}\mathbf{x}$ and $\mathbf{x} = \mathbf{F}^{H} \pmb{\hat{x}}$. Applying both $\mathbf{F}^{H}$ and $\mathbf{F}$ we get
$$\mathbf{F}^{H} \mathbf{F} \mathbf{x} = \mathbf{F}^{H} \pmb{\hat{x}} = \mathbf{x} \implies \mathbf{F}^{H} \mathbf{F} = \pmb{I}.$$
Thus, the DFT matrix $\mathbf{F}$ is unitary.

\subsection{Extended Derivations}
Here we provide details on obtaining $\mathbf{\hat{M}}$ and $\mathbf{\hat{a}}$ as described by Equation \ref{eq:3.8} of the main text. This enables us to update $\mathbf{u}$ efficiently, as mentioned in Section 4.2 of the main text. 

We begin with the following equation
\begin{equation}
        \mathbf{M}\mathbf{\tilde{u}^{(k+1)}} = \mathbf{a} \label{eq:A1},
\end{equation}
where $\mathbf{M} = \Big((\mathbf{\tilde{F}_{P}^{-1}})^{H}\mathbf{\tilde{F}_{P}^{-1}} + \beta \mathbf{\tilde{I}} \Big) \in \mathbb{C}^{N^{2} \times N^{2}}$ and $\mathbf{a} = \Big((\mathbf{\tilde{F}_{P}^{-1}})^H \mathbf{\tilde{b}} + \beta\Big(\mathbf{\tilde{z}^{(k)}} - \frac{\mathbf{\tilde{y}^{(k)}}}{\beta}\Big)\Big) \in \mathbb{C}^{N^{2}}$.
Multiplying both sides of Equation \ref{eq:A1} by $\mathbf{\tilde{F}^{-1}} = \mathbf{F^{-1}} \bigotimes \mathbf{F^{-1}} \in \mathbb{C}^{N^{2} \times N^{2}}$ and simplifying using the properties of the discrete Fourier transform matrix, the left-hand side of the equation becomes:
\begin{align*}
    \mathbf{\tilde{F}^{-1} M \tilde{u}^{(k+1)}} &= \mathbf{\tilde{F}^{-1}}\Big((\mathbf{\tilde{F}_{P}^{-1}})^{H}\mathbf{\tilde{F}_{P}^{-1}} + \beta \mathbf{\tilde{I}} \Big)\mathbf{\tilde{u}^{(k+1)}}\\
    &=\mathbf{\tilde{F}^{-1}}(\mathbf{(\tilde{P}\tilde{F}^{-1}})^H \mathbf{\tilde{P}\tilde{F}^{-1}} + \beta \mathbf{\tilde{I}})\mathbf{\tilde{u}^{(k+1)}} \quad [\mathbf{\tilde{F}^{-1}} = \mathbf{\tilde{F}}^{H} \textrm{ because DFT matrix is unitary}]\\
    &= \mathbf{\tilde{F}^{-1}}((\mathbf{\tilde{F}}^{H})^H \mathbf{\tilde{P}}^{H}\mathbf{\tilde{P}} \mathbf{\tilde{F}^{-1}} + \beta \mathbf{\tilde{I}})\mathbf{\tilde{u}^{(k+1)}} \quad [\mathbf{\tilde{P}}^{H} = \mathbf{\tilde{P}}^{T} \textrm{ because entries of $\mathbf{\tilde{P}}$ are real by construction}]\\
    &= \mathbf{\tilde{F}^{-1}}\mathbf{\tilde{F}\tilde{P}}^{T}\mathbf{\tilde{P}\tilde{F}^{-1}}\mathbf{\tilde{u}^{(k+1)}} + \beta\mathbf{\tilde{F}^{-1}}\mathbf{\tilde{u}^{(k+1)}} \quad [\beta \in \mathbb{R} \textrm{ is a constant}]\\
    &= \mathbf{\tilde{P}}^{T}\mathbf{\tilde{P}\tilde{F}^{-1}}\mathbf{\tilde{u}^{(k+1)}} + \beta\mathbf{\tilde{F}^{-1}}\mathbf{\tilde{u}^{(k+1)}} \\
    &= ( \mathbf{\tilde{P}}^{T}\mathbf{\tilde{P}} + \beta\mathbf{\tilde{I}})\mathbf{\tilde{F}^{-1}\tilde{u}^{(k+1)}}\\
    &= \mathbf{\hat{M} \tilde{F}^{-1}\tilde{u}^{(k+1)}}
\end{align*}

Similarly, the right-hand side of Equation \ref{eq:A1} becomes:
\begin{align*}
    \mathbf{\tilde{F}^{-1}a} &= \mathbf{\tilde{F}^{-1}}\Big[(\mathbf{\tilde{F}_{P}^{-1}})^H \mathbf{\tilde{b}} \mathbf{\tilde{F}_{P}^{-1}} + \beta\Big(\mathbf{\tilde{z}^{(k)}} - \frac{\mathbf{\tilde{y}^{(k)}}}{\beta}\Big)\Big] \\
    &= \mathbf{\tilde{F}^{-1}}\Big[(\mathbf{\tilde{P}\tilde{F}^{-1}})^H \mathbf{\tilde{b}} \mathbf{\tilde{P}\tilde{F}^{-1}} + \beta\Big(\mathbf{\tilde{z}^{(k)}} - \frac{\mathbf{\tilde{y}^{(k)}}}{\beta}\Big)\Big] \\
    &= \mathbf{\tilde{F}^{-1}}(\mathbf{\tilde{F}}^{H})^H \mathbf{\tilde{P}}^{H} \mathbf{\tilde{b}} \mathbf{\tilde{P}\tilde{F}^{-1}} + \beta \mathbf{\tilde{F}^{-1} \tilde{z}^{(k)}} - \mathbf{\tilde{F}^{-1} \tilde{y}^{(k)}} \\
    &= \mathbf{\tilde{F}^{-1}\tilde{F} \tilde{P}}^{T} \mathbf{\tilde{b}} + \beta \mathbf{\tilde{F}^{-1}\tilde{z}^{(k)}} \\
    &=  \mathbf{\tilde{P}}^{T} \mathbf{\tilde{b}} + \beta \mathbf{\tilde{F}^{-1} \tilde{z}^{(k)}} \\
    &= \mathbf{\hat{a}}
    \hspace{20em} \qed
\end{align*}

\subsection{Proofs of propositions}
We present the proofs of propositions \ref{prop:2} and \ref{prop:3} described in Section 4.2 of the main text, first recalling the statement of results:

\begin{proposition}
$\!\mathbf{4.1}$
The functions $f(\mathbf{U}) = \frac{1}{2}||\mathbf{F_{P}^{-1}}\mathbf{U}(\mathbf{F_{P}^{-1}})^{H} - \mathbf{B}||^2_2$ and $g(\mathbf{Z}) = \lambda||\mathbf{Z}||_1$ are closed, convex, and proper. 
\end{proposition}

The proof of Proposition \ref{prop:1} has been provided in the main text. Proposition $\mathbf{4.1}$ implies that there exist $\mathbf{U}$ and $\mathbf{Z}$, not necessarily unique, that minimize the augmented Lagrangian, thereby making the $\mathbf{U}$ and $\mathbf{Z}$ updates
\begin{equation}
\begin{split}
    \mathbf{U^{(k+1)}} &:= \argmin_\mathbf{U} L_{\beta}(\mathbf{U}, \mathbf{Z^{(k)}}, \mathbf{Y^{(k)}}),\\
    \mathbf{Z^{(k+1)}} &:= \argmin_\mathbf{Z} L_{\beta}(\mathbf{U^{(k+1)}}, \mathbf{Z}, \mathbf{Y^{(k)}}),\\
\end{split}
\end{equation}
solvable. 

\begin{proposition}
$\!\mathbf{4.2}$
For $\beta=0$, $L_{0} = \frac{1}{2}||\mathbf{F_{P}^{-1}U}(\mathbf{F_{P}^{-1}})^{H} - \mathbf{B}||_{2}^2 + \lambda||\mathbf{Z}||_1 +  \mathbf{Y}^{T}(\mathbf{U}-\mathbf{Z})$, the unaugmented Lagrangian has a saddle point for all $\mathbf{U},\mathbf{Z},\mathbf{Y}$. 
\end{proposition}

\begin{proof}
A Lagrangian $L_{0}$ is said to have a saddle point when there exist $(\mathbf{U^{*}},\mathbf{Z^{*}},\mathbf{Y^{*}})$ not necessarily unique such that,
$$L_{0}(\mathbf{U^{*},Z^{*},Y}) \leq L_{0}(\mathbf{U^{*},Z^{*},Y^{*}}) \leq L_{0}(\mathbf{U,Z,Y^{*}}).$$
From Proposition $\mathbf{4.1}$ it follows that $L_{0}(\mathbf{U^{*},Z^{*},Y^{*}})$ is finite for any saddle
point $(\mathbf{U^{*},Z^{*},Y^{*}})$. This implies not only that $(\mathbf{U^{*},Z^{*}})$ is a solution to the primal objective function $\frac{1}{2}||\mathbf{F_{P}^{-1}U}(\mathbf{F_{P}^{-1}})^{H} - \mathbf{B}||_{2}^2 + \lambda||\mathbf{Z}||_1$, thus $\mathbf{U^{*}} = \mathbf{Z^{*}}$ and $\frac{1}{2}||\mathbf{F_{P}^{-1}U^{*}}(\mathbf{F_{P}^{-1}})^{H} - \mathbf{B}||_{2}^2 < \infty$ and $\lambda||\mathbf{Z^{*}}||_1 < \infty$, but also that strong duality holds and $\mathbf{Y^{*}}$ is a dual optimal, i.e. it maximizes the dual function $\mathbf{Y}^{T}(\mathbf{U}-\mathbf{Z})$ \citep{boyd2004convex}.

We can easily verify that
\begin{equation*}
    \frac{1}{2}||\mathbf{F_{P}^{-1}U^{*}}(\mathbf{F_{P}^{-1}})^{H} - \mathbf{B}||_{2}^2 + \lambda||\mathbf{U}^{*}||_1 = \frac{1}{2}||\mathbf{F_{P}^{-1}U^{*}}(\mathbf{F_{P}^{-1}})^{H} - \mathbf{B}||_{2}^2 + \lambda||\mathbf{U}^{*}||_1 < \frac{1}{2}||\mathbf{F_{P}^{-1}U}(\mathbf{F_{P}^{-1}})^{H} - \mathbf{B}||_{2}^2 + \lambda||\mathbf{Z}||_1 + (\mathbf{Y^{*}})^{T}(\mathbf{U}-\mathbf{Z}),
\end{equation*}
where the last inequality follows from $(\mathbf{U^{*},Z^{*}})$ minimizing $\frac{1}{2}||\mathbf{F_{P}^{-1}U}(\mathbf{F_{P}^{-1}})^{H} - \mathbf{B}||_{2}^2 + \lambda||\mathbf{Z}||_1$ among all $(\mathbf{U},\mathbf{Z})$ and $\mathbf{Y^{*}}$ maximizing the dual objective function among all $\mathbf{Y}$. 

\end{proof}

\begin{proposition}
$\!\mathbf{4.3}$
    Under propositions $\mathbf{4.1}$ and $\mathbf{4.2}$, the ADMM algorithm achieves the following:
    \begin{enumerate}
        \item Primal residual convergence: $\mathbf{R^{(k+1)}} = \mathbf{U^{(k+1)}} - \mathbf{Z^{(k+1)}} \to 0$  as $k \to \infty$.
        \item Dual residual convergence: $\mathbf{S^{(k+1)}} = \beta(\mathbf{Z^{(k+1)}} - \mathbf{Z^{(k)}}) \to 0$ as $k \to \infty.$
    \end{enumerate}
\end{proposition}

\begin{proof}
The proof of proposition $\mathbf{4.3}$ \citep{gabay1983chapter, eckstein1992douglas, boyd2011distributed} is divided into three parts based on three central inequalities, all of which we prove as part of this proof.

Let $(\mathbf{U^{*}},\mathbf{Z^{*}},\mathbf{Y^{*}})$ be a saddle point for the unaugmented Lagrangian $L_{0} = \frac{1}{2}||\mathbf{F_{P}^{-1}U}(\mathbf{F_{P}^{-1}})^{H} - \mathbf{B}||_{2}^2 + \lambda||\mathbf{Z}||_1 +  \mathbf{Y}^{T}(\mathbf{U}-\mathbf{Z})$. Equivalently let $(\mathbf{\tilde{u}^{*}},\mathbf{\tilde{z}^{*}},\mathbf{\tilde{y}^{*}})$ be the saddle point for the unaugmented Lagrangian $L_{0} = \frac{1}{2}||\mathbf{\tilde{F}_{P}^{-1}} \mathbf{\tilde{u}} - \mathbf{\tilde{b}}||_{2}^2 + \lambda||\mathbf{\tilde{z}}||_1 +  \mathbf{\tilde{y}}^{T}(\mathbf{\tilde{u}}-\mathbf{\tilde{z}})$ and consider the first inequality
\begin{equation}
    q^{*} - q^{(k+1)} \leq (\mathbf{\tilde{y}}^{*})^{T}\mathbf{\tilde{r}^{(k+1)}},
    \label{ineq:1}
\end{equation}
where $q^{*}$ is the optimal value to which the objective function converges as $k \to \infty$.

\subsubsection{Proof of Inequality \ref{ineq:1}}
The following inequality holds
\begin{equation}
    L_{0}(\mathbf{\tilde{u}^{*},\tilde{z}^{*},\tilde{y}^{*}}) \leq L_{0}(\mathbf{\tilde{u}^{(k+1)},\tilde{z}^{(k+1)},\tilde{y}^{*}})
    \label{ineq:2}
\end{equation}
because $(\mathbf{\tilde{u}^{*}},\mathbf{\tilde{z}^{*}},\mathbf{\tilde{y}^{*}})$ is a saddle point for $L_{0}$. Using the constraint of the optimization problem $\mathbf{\tilde{u}^{*}}=\mathbf{\tilde{z}^{*}}$, the left-hand side of inequality \ref{ineq:2} reduces to $q^{*}$. 
The right-hand side of inequality \ref{ineq:2} can be simplified as
\begin{align*}
    L_{0}(\mathbf{\tilde{u}^{(k+1)}, \tilde{z}^{(k+1)}, \tilde{y}^{*}}) &= \frac{1}{2}||\mathbf{\tilde{F}_{P}^{-1}\tilde{u}^{(k+1)}} - \mathbf{\tilde{b}}||_{2}^2 + \lambda||\mathbf{\tilde{z}^{(k+1)}}||_1 +  (\mathbf{\tilde{y}}^{*})^{T}(\mathbf{\tilde{u}^{(k+1)}}-\mathbf{\tilde{z}^{(k+1)}})\\
    &= q^{(k+1)} + (\mathbf{\tilde{y}}^{*})^{T} \mathbf{\tilde{r}^{(k+1)}} \quad \Big[\mathbf{\tilde{r}^{(k+1)}} = \mathbf{\tilde{u}^{(k+1)}}-\mathbf{\tilde{z}^{(k+1)}} ; q^{(k+1)} = \frac{1}{2}||\mathbf{\tilde{F}_{P}^{-1}\tilde{u}^{(k+1)}} - \mathbf{\tilde{b}}||_{2}^2 + \lambda||\mathbf{\tilde{z}^{(k+1)}}||_1 \Big]. \\
\end{align*}
Thus, we have 
\begin{align*}
    q^{*} &\leq q^{(k+1)} + (\mathbf{\tilde{y}}^{*})^{T} \mathbf{\tilde{r}^{(k+1)}}\\
    q^{*} - q^{(k+1)} &\leq (\mathbf{\tilde{y}}^{*})^{T} \mathbf{\tilde{r}^{(k+1)}}.
    \hspace{10em}\qed
\end{align*}

Next, we consider the second inequality  

\begin{equation}
    q^{(k+1)} - q^{*} \leq -\mathbf{(\tilde{y}^{(k+1)}})^{T} \mathbf{\tilde{r}^{(k+1)}} - \beta(\mathbf{\tilde{z}^{(k+1)}} - \mathbf{\tilde{z}^{(k)}})^{T}(\mathbf{\tilde{r}^{(k+1)}} + (\mathbf{\tilde{z}^{(k+1)}} - \mathbf{\tilde{z}^{*}})).
    \label{ineq:3}
\end{equation}
\subsubsection{Proof of Inequality \ref{ineq:3}}
We know that, by definition, $\mathbf{\tilde{u}^{(k+1)}}$ minimizes $L_{\beta}(\mathbf{\tilde{u}}, \mathbf{\tilde{z}^{(k)}}, \mathbf{\tilde{y}^{(k)}})$. Since $f(\mathbf{\tilde{u}}) = \frac{1}{2}||\mathbf{\tilde{F}_{P}^{-1}}\mathbf{\tilde{u}} - \mathbf{\tilde{b}}||^2_2$ is closed, proper, and convex by Proposition $\mathbf{4.1}$, it is differentiable, implying that $L_{\beta}$ is subdifferentiable. The necessary and sufficient optimality condition is

\begin{align}
    0 &\in \partial{L_{\beta}}(\mathbf{\tilde{u}^{(k+1)}}, \mathbf{\tilde{z}^{(k)}}, \mathbf{\tilde{y}^{(k)}}) = \partial{\Big(\frac{1}{2}||\mathbf{\tilde{F}_P^{-1}\tilde{u}^{(k+1)}} - \mathbf{\tilde{b}}||_{2}^2 +
        \lambda||\mathbf{\tilde{z}^{(k)}}||_1 +   (\mathbf{\tilde{y}^{(k)}})^{T}(\mathbf{\tilde{u}^{(k+1)}}-\mathbf{\tilde{z}^{(k)}}) + \frac{\beta}{2}||\mathbf{\tilde{u}^{(k+1)}}-\mathbf{\tilde{z}^{(k)}}||_{2}^2\Big)} \nonumber\\
        &= \partial{\Big(\frac{1}{2}||\mathbf{\tilde{F}_P^{-1}\tilde{u}^{(k+1)}} - \mathbf{\tilde{b}}||_{2}^2 \Big)} + \partial{(\lambda||\mathbf{\tilde{z}^{(k)}}||_1}) + 
        \partial{((\mathbf{\tilde{y}^{(k)}})^{T}(\mathbf{\tilde{u}^{(k+1)}}-\mathbf{\tilde{z}^{(k)}}))} + \partial{\Big(\frac{\beta}{2}||\mathbf{\tilde{u}^{(k+1)}}-\mathbf{\tilde{z}^{(k)}}||_{2}^2\Big)} \nonumber\\
        &= \partial{\Big(\frac{1}{2}||\mathbf{\tilde{F}_P^{-1}\tilde{u}^{(k+1)}} - \mathbf{\tilde{b}}||_{2}^2 \Big)} + \mathbf{\tilde{y}^{(k)}}
        + \beta(\mathbf{\tilde{u}^{(k+1)}} - \mathbf{\tilde{z}^{(k)}}) \label{eq:1},
\end{align}
where we use the fact that the subdifferential of the sum of a subdifferential function and a differentiable function with domain $\mathbb{R}^{N}$ is the sum of the subdifferential and the gradient \citep{rockafellar1970convex}. 

Since the update of $\mathbf{\tilde{y}}$ is
$$\mathbf{\tilde{y}^{(k+1)}} = \mathbf{\tilde{y}^{(k)}} + \beta(\mathbf{\tilde{u}^{(k+1)}} - \mathbf{\tilde{z}^{(k+1)}}),$$
we can substitute $\mathbf{\tilde{y}^{(k)}} = \mathbf{\tilde{y}^{(k+1)}} - \beta(\mathbf{\tilde{u}^{(k+1)}} - \mathbf{\tilde{z}^{(k+1)}})$ in \ref{eq:1} and rearrange the terms to obtain

\begin{align*}
  0 &\in \partial{\Big(\frac{1}{2}||\mathbf{\tilde{F}_P^{-1}\tilde{u}^{(k+1)}} - \mathbf{\tilde{b}}||_{2}^2 \Big)}  + \mathbf{\tilde{y}^{(k+1)}} - \beta(\mathbf{\tilde{u}^{(k+1)}} - \mathbf{\tilde{z}^{(k+1)}}) + \beta \mathbf{\tilde{u}^{(k+1)}} - \beta\mathbf{\tilde{z}^{(k)}} \\
  &= \partial{\Big(\frac{1}{2}||\mathbf{\tilde{F}_P^{-1}\tilde{u}^{(k+1)}} - \mathbf{\tilde{b}}||_{2}^2 \Big)}  + \mathbf{\tilde{y}^{(k+1)}} + \beta(\mathbf{\tilde{z}^{(k+1)}} - \mathbf{\tilde{z}^{(k)}}).
\end{align*}

This implies that $\mathbf{\tilde{u}^{(k+1)}}$ minimizes 
$$\frac{1}{2}||\mathbf{\tilde{F}_P^{-1}\tilde{u}^{(k+1)}} - \mathbf{\tilde{b}}||_{2}^2 + (\mathbf{\tilde{y}^{(k+1)}} + \beta(\mathbf{\tilde{z}^{(k+1)}} - \mathbf{\tilde{z}^{(k)}}))^{T} \mathbf{\tilde{u}}.$$
Analogously we can show that $\mathbf{\tilde{z}^{(k+1)}}$ minimizes $\lambda||\mathbf{\tilde{z}}||_1 - (\mathbf{\tilde{y}^{(k+1)}})^{T}\mathbf{\tilde{z}}$ and it follows that,
\begin{align}
    \frac{1}{2}||\mathbf{\tilde{F}_P^{-1}\tilde{u}^{(k+1)}} - \mathbf{\tilde{b}}||_{2}^2 + (\mathbf{\tilde{y}^{(k+1)}} + \beta(\mathbf{\tilde{z}^{(k+1)}} - \mathbf{\tilde{z}^{(k)}}))^{T} \mathbf{\tilde{u}^{(k+1)}}\nonumber\\
    \leq \frac{1}{2}||\mathbf{\tilde{F}_P^{-1}\tilde{u}^{*}} - \mathbf{\tilde{b}}||_{2}^2 + (\mathbf{\tilde{y}^{(k+1)}} + \beta(\mathbf{\tilde{z}^{(k+1)}} - \mathbf{\tilde{z}^{(k)}}))^{T} \mathbf{\tilde{u}^{*}} \label{ineq:4}
\end{align}
and 
\begin{equation}
\lambda||\mathbf{\tilde{z}^{(k+1)}}||_1 - (\mathbf{\tilde{y}^{(k+1)}})^{T}\mathbf{\tilde{z}^{(k+1)}} \leq \lambda||\mathbf{\tilde{z}^{*}}||_1 - (\mathbf{\tilde{y}^{(k+1)}})^{T}\mathbf{\tilde{z}^{*}}
\label{ineq:5}
\end{equation}

Adding the inequalities \ref{ineq:4} and \ref{ineq:5}, we obtain
\begin{align}
     &\frac{1}{2}||\mathbf{\tilde{F}_P^{-1}\tilde{u}^{(k+1)}} - \mathbf{\tilde{b}}||_{2}^2 + (\mathbf{\tilde{y}^{(k+1)}} + \beta(\mathbf{\tilde{z}^{(k+1)}} - \mathbf{\tilde{z}^{(k)}}))^{T} \mathbf{\tilde{u}^{(k+1)}} + \lambda||\mathbf{\tilde{z}^{(k+1)}}||_1 - (\mathbf{\tilde{y}^{(k+1)}})^{T}\mathbf{\tilde{z}^{(k+1)}} \nonumber\\
     &\leq \frac{1}{2}||\mathbf{\tilde{F}_P^{-1}\tilde{u}^{*}} - \mathbf{\tilde{b}}||_{2}^2 + (\mathbf{\tilde{y}^{(k+1)}} + \beta(\mathbf{\tilde{z}^{(k+1)}} - \mathbf{\tilde{z}^{(k)}}))^{T} \mathbf{\tilde{u}^{*}} + \lambda||\mathbf{\tilde{z}^{*}}||_1 - (\mathbf{\tilde{y}^{(k+1)}})^{T}\mathbf{\tilde{z}^{*}}.
     \label{ineq:6}
\end{align}

We can rearrange and simplify the terms on the left-hand side of inequality \ref{ineq:6},

\begin{align}
     &\frac{1}{2}||\mathbf{\tilde{F}_P^{-1}u^{(k+1)}} - \mathbf{\tilde{b}}||_{2}^2 + (\mathbf{\tilde{y}^{(k+1)}} + \beta(\mathbf{\tilde{z}^{(k+1)}} - \mathbf{\tilde{z}^{(k)}}))^{T} \mathbf{\tilde{u}^{(k+1)}} + \lambda||\mathbf{\tilde{z}^{(k+1)}}||_1 - (\mathbf{\tilde{y}^{(k+1)}})^{T}\mathbf{\tilde{z}^{(k+1)}} \nonumber\\
     &= \frac{1}{2}||\mathbf{\tilde{F}_P^{-1}\tilde{u}^{(k+1)}} - \mathbf{\tilde{b}}||_{2}^2 + \lambda||\mathbf{\tilde{z}^{(k+1)}}||_1 + (\mathbf{\tilde{y}^{(k+1)}})^{T} (\mathbf{\tilde{u}^{(k+1)}} - \mathbf{\tilde{z}^{(k+1)}}) +  \beta(\mathbf{\tilde{z}^{(k+1)}} - \mathbf{\tilde{z}^{(k)}})^{T} \mathbf{\tilde{u}^{(k+1)}} \quad [\textrm{Rearranging the terms}] \nonumber\\
     &= q^{(k+1)} + (\mathbf{\tilde{y}^{(k+1)}})^{T} \mathbf{\tilde{r}^{(k+1)}} +  \beta(\mathbf{\tilde{z}^{(k+1)}} - \mathbf{\tilde{z}^{(k)}})^{T} (\mathbf{\tilde{r}^{(k+1)}} + \mathbf{\tilde{z}^{(k+1)}}) \quad [\textrm{Because }\mathbf{\tilde{r}^{(k+1)}} = \mathbf{\tilde{u}^{(k+1)}} - \mathbf{\tilde{z}^{(k+1)}} ] \label{eq:2}
\end{align}

Similarly, we can also simplify the terms on the right-hand side of the inequality \ref{ineq:6},

\begin{align}
    &\frac{1}{2}||\mathbf{\tilde{F}_P^{-1}\tilde{u}^{*}} - \mathbf{\tilde{b}}||_{2}^2 + (\mathbf{\tilde{y}^{(k+1)}} + \beta(\mathbf{\tilde{z}^{(k+1)}} - \mathbf{\tilde{z}^{(k)}}))^{T} \mathbf{\tilde{u}^{*}} + \lambda||\mathbf{\tilde{z}^{*}}||_1 - (\mathbf{\tilde{y}^{(k+1)}})^{T}\mathbf{\tilde{z}^{*}} \nonumber\\
     &= \frac{1}{2}||\mathbf{\tilde{F}_P^{-1}\tilde{u}^{*}} - \mathbf{\tilde{b}}||_{2}^2 + \lambda||\mathbf{\tilde{z}^{*}}||_1 + (\mathbf{\tilde{y}^{(k+1)}})^{T} (\mathbf{\tilde{u}^{*}} - \mathbf{\tilde{z}^{*}}) +  \beta(\mathbf{\tilde{z}^{(k+1)}} - \mathbf{\tilde{z}^{(k)}})^{T} \mathbf{\tilde{u}^{*}} \quad [\textrm{Rearranging the terms}] \nonumber\\
    &= q^{*} + \beta(\mathbf{\tilde{z}^{(k+1)}} - \mathbf{\tilde{z}^{(k)}})^{T} \mathbf{\tilde{z}^{*}} \quad [\textrm{Because of the optimization constraint } \mathbf{\tilde{u}^{*}} = \mathbf{\tilde{z}^{*}} ] 
    \label{eq:3}
\end{align}

We can combine \ref{eq:2} and \ref{eq:3} and rearrange the terms to obtain inequality \ref{ineq:3}

\begin{align*}
    q^{(k+1)} - q^{*} \leq -\mathbf{(\tilde{y}^{(k+1)}})^{T} \mathbf{\tilde{r}^{(k+1)}} - \beta(\mathbf{\tilde{z}^{(k+1)}} - \mathbf{\tilde{z}^{(k)}})^{T}(\mathbf{\tilde{r}^{(k+1)}} + (\mathbf{\tilde{z}^{(k+1)}} - \mathbf{\tilde{z}^{*}})).
    \qed
\end{align*}


Now that we have proved the inequalities \ref{ineq:2} and \ref{ineq:3}), only the last inequality

\begin{equation}
    V^{(k)} - V^{(k+1)} \geq  \beta ||\mathbf{\tilde{r}^{(k+1)}}||_{2}^{2} + \beta||\mathbf{(\tilde{z}^{(k+1)}} - \mathbf{\tilde{z}^{(k)}})||_{2}^{2},
    \label{ineq:7}
\end{equation}
remains to be proven. 

\subsubsection{Proof of Inequality \ref{ineq:7}}
Recall that $(\mathbf{\tilde{u}^{*}},\mathbf{\tilde{z}^{*}},\mathbf{\tilde{y}^{*}})$ is a saddle point for the unaugmented Lagrangian $L_{0}$ and define
$$V^{(k)} := \frac{1}{\beta} ||\mathbf{\tilde{y}^{(k)}} - \mathbf{\tilde{y}}^{*}||_{2}^{2} + \beta||\mathbf{(\tilde{z}^{(k)}} - \mathbf{\tilde{z}^{*})}||_{2}^{2}.$$

Adding inequalities \ref{ineq:2} and \ref{ineq:3}, rearranging the terms, and multiplying by 2 on both sides gives us

\begin{align}
    2(\mathbf{\tilde{y}^{(k+1)}} - \mathbf{\tilde{y}^{*}})^{T}\mathbf{\tilde{r}^{(k+1)}} + 2\beta(\mathbf{\tilde{z}^{(k+1)}} - \mathbf{\tilde{z}^{(k)}})^{T}\mathbf{\tilde{r}^{(k+1)}} + 2\beta(\mathbf{\tilde{z}^{(k+1)}} - \mathbf{\tilde{z}^{(k)}})^{T}(\mathbf{\tilde{z}^{(k+1)}} - \mathbf{\tilde{z}^{*}}) \leq 0.
    \label{ineq:8}
\end{align}

The first term of inequality \ref{ineq:8} can be rewritten as

\begin{align*}
    2(\mathbf{\tilde{y}^{(k+1)}} - \mathbf{\tilde{y}^{*}})^{T}\mathbf{\tilde{r}^{(k+1)}} &= 2(\mathbf{\tilde{y}^{(k)}} + \beta\mathbf{\tilde{r}^{(k+1)}} - \mathbf{\tilde{y}^{*}})^{T}\mathbf{\tilde{r}^{(k+1)}} \qquad [\textrm{Because } \mathbf{\tilde{y}^{(k+1)}} = \mathbf{\tilde{y}^{(k)}} + \beta\mathbf{\tilde{r}^{(k+1)}}]\\
    &= 2(\mathbf{\tilde{y}^{(k)}} - \mathbf{\tilde{y}^{*}})^{T}\mathbf{\tilde{r}^{(k+1)}} + 2\beta(\mathbf{\tilde{r}^{(k+1)}})^{T}\mathbf{\tilde{r}^{(k+1)}} \qquad [\beta^{T} = \beta \textrm{ since } \beta \in \mathbb{R}]\\
    &= 2(\mathbf{\tilde{y}^{(k)}} - \mathbf{\tilde{y}^{*}})^{T}\mathbf{\tilde{r}^{(k+1)}} + \beta||\mathbf{\tilde{r}^{(k+1)}}||_{2}^{2} + \beta||\mathbf{\tilde{r}^{(k+1)}}||_{2}^{2}, \qquad[(\mathbf{\tilde{r}^{(k+1)}})^{T}\mathbf{\tilde{r}^{(k+1)}} = ||\mathbf{\tilde{r}^{(k+1)}}||_{2}^{2}]
\end{align*}

and plugging $\mathbf{\tilde{r}^{(k+1)}} = \frac{1}{\beta}(\mathbf{\tilde{y}^{(k+1)}} - \mathbf{\tilde{y}^{(k)}})$ in the first two terms gives

\begin{align}
    &\frac{2}{\beta}(\mathbf{\tilde{y}^{(k)}} - \mathbf{\tilde{y}^{*}})^{T}(\mathbf{\tilde{y}^{(k+1)}} - \mathbf{\tilde{y}^{(k)}}) + \frac{1}{\beta} ||\mathbf{\tilde{y}^{(k+1)}} - \mathbf{\tilde{y}^{(k)}}||_{2}^{2} + \beta||\mathbf{\tilde{r}^{(k+1)}}||_{2}^{2} \nonumber\\
    &= \frac{1}{\beta}\Big(||\mathbf{\tilde{y}^{(k+1)}} - \mathbf{\tilde{y}^{*}}||_{2}^{2} - ||\mathbf{\tilde{y}^{(k)}} - \mathbf{\tilde{y}^{*}}||_{2}^{2}\Big) + \beta||\mathbf{\tilde{r}^{(k+1)}}||_{2}^{2} \qquad[\textrm{Adding and subtracting } \mathbf{\tilde{y}^{*}}].\label{eq:4}
\end{align}

We can now rewrite the remaining terms of the inequality \ref{ineq:8}

\begin{align}
    &\beta||\mathbf{\tilde{r}^{(k+1)}}||_{2}^{2} + 2\beta(\mathbf{\tilde{z}^{(k+1)}} - \mathbf{\tilde{z}^{(k)}})^{T}\mathbf{\tilde{r}^{(k+1)}} + 2\beta(\mathbf{\tilde{z}^{(k+1)}} - \mathbf{\tilde{z}^{(k)}})^{T}(\mathbf{\tilde{z}^{(k+1)}} - \mathbf{\tilde{z}^{*}}),
    \label{eq:5}
\end{align}

where the term $\beta||\mathbf{\tilde{r}^{(k+1)}}||_{2}^{2}$ is extracted from Equation \ref{eq:4}. Adding and subtracting $\mathbf{\tilde{z}^{(k)}}$ from the last term of \ref{eq:5} gives

\begin{align*}
    \beta||\mathbf{\tilde{r}^{(k+1)}} + (\mathbf{\tilde{z}^{(k+1)}} - \mathbf{\tilde{z}^{(k)}})||_{2}^{2} + \beta||(\mathbf{\tilde{z}^{(k+1)}} - \mathbf{\tilde{z}^{(k)}})||_{2}^{2} + 2\beta(\mathbf{\tilde{z}^{(k+1)}} - \mathbf{\tilde{z}^{(k)}})^{T}(\mathbf{\tilde{z}^{(k)}} - \mathbf{\tilde{z}^{*}}),
\end{align*}

and by substituting $\mathbf{\tilde{z}^{(k+1)}} - \mathbf{\tilde{z}^{(k)}} = (\mathbf{\tilde{z}^{(k+1)}} - \mathbf{\tilde{z}^{*}}) - (\mathbf{\tilde{z}^{(k)}} - \mathbf{\tilde{z}^{*}})$ in the last two terms, we get

\begin{align*}
\beta||\mathbf{\tilde{r}^{(k+1)}} + (\mathbf{\tilde{z}^{(k+1)}} - \mathbf{\tilde{z}^{(k)}})||_{2}^{2} + \beta\Big(||\mathbf{\tilde{z}^{(k+1)}} - \mathbf{\tilde{z}^{*}}||_{2}^{2} - ||\mathbf{\tilde{z}^{(k)}} - \mathbf{\tilde{z}^{*}}||_{2}^{2}\Big).
\end{align*}

This implies that inequality \ref{ineq:8} can be written as

\begin{align*}
 \frac{1}{\beta}\Big(||\mathbf{\tilde{y}^{(k+1)}} - \mathbf{\tilde{y}^{*}}||_{2}^{2} - ||\mathbf{\tilde{y}^{(k)}} - \mathbf{\tilde{y}^{*}}||_{2}^{2}\Big) +\beta||\mathbf{\tilde{r}^{(k+1)}} + (\mathbf{\tilde{z}^{(k+1)}} - \mathbf{\tilde{z}^{(k)}})||_{2}^{2} + \beta\Big(||\mathbf{\tilde{z}^{(k+1)}} - \mathbf{\tilde{z}^{*}}||_{2}^{2} - ||\mathbf{\tilde{z}^{(k)}} - \mathbf{\tilde{z}^{*}}||_{2}^{2}\Big) \leq 0 \\
 \implies V^{(k)} - V^{(k+1)} \geq \beta||\mathbf{\tilde{r}^{(k+1)}} + (\mathbf{\tilde{z}^{(k+1)}} - \mathbf{\tilde{z}^{(k)}})||_{2}^{2}
\end{align*}

We can expand $\beta||\mathbf{\tilde{r}^{(k+1)}} + (\mathbf{\tilde{z}^{(k+1)}} - \mathbf{\tilde{z}^{(k)}})||_{2}^{2}$ as
\begin{align*}
\beta||\mathbf{\tilde{r}^{(k+1)}}||_{2}^{2} + 2\beta(\mathbf{\tilde{r}^{(k+1)}})^{T}(\mathbf{\tilde{z}^{(k+1)}} - \mathbf{\tilde{z}^{(k)}}) + \beta||(\mathbf{\tilde{z}^{(k+1)}} - \mathbf{\tilde{z}^{(k)}})||_{2}^{2}.\\
\end{align*}

Thus, to prove the third inequality \ref{ineq:7}, it suffices to show that $2\beta(\mathbf{\tilde{r}^{(k+1)}})^{T}(\mathbf{\tilde{z}^{(k+1)}} - \mathbf{\tilde{z}^{(k)}})$ is positive. Recall that $\mathbf{\tilde{z}^{(k+1)}}$ minimizes $\lambda||\mathbf{\tilde{z}}||_1 - (\mathbf{\tilde{y}^{(k+1)}})^{T}\mathbf{\tilde{z}}$ and similarly, $\mathbf{\tilde{z}^{(k)}}$ minimizes 
$\lambda||\mathbf{\tilde{z}}||_1 - (\mathbf{\tilde{y}^{(k)}})^{T}\mathbf{\tilde{z}}$ which allows us to add

$$\lambda||\mathbf{\tilde{z}^{(k+1)}}||_1 - (\mathbf{\tilde{y}^{(k+1)}})^{T}\mathbf{\tilde{z}^{k+1}} \leq \lambda||\mathbf{\tilde{z}^{(k)}}||_1 - (\mathbf{\tilde{y}^{(k+1)}})^{T}\mathbf{\tilde{z}^{k}}$$
and 
$$\lambda||\mathbf{\tilde{z}^{(k)}}||_1 - (\mathbf{\tilde{y}^{(k)}})^{T}\mathbf{\tilde{z}^{k}} \leq \lambda||\mathbf{\tilde{z}^{(k+1)}}||_1 - (\mathbf{\tilde{y}^{(k)}})^{T}\mathbf{\tilde{z}^{k+1}}$$
to get
\begin{align*}
    \lambda||\mathbf{\tilde{z}^{(k+1)}}||_1 - (\mathbf{\tilde{y}^{(k+1)}})^{T}\mathbf{\tilde{z}^{k+1}} + \lambda||\mathbf{\tilde{z}^{(k)}}||_1 - (\mathbf{\tilde{y}^{(k)}})^{T}\mathbf{\tilde{z}^{k}}\\
    &\leq \lambda||\mathbf{\tilde{z}^{(k)}}||_1 - (\mathbf{\tilde{y}^{(k+1)}})^{T}\mathbf{\tilde{z}^{k}} + \lambda||\mathbf{\tilde{z}^{(k+1)}}||_1 - (\mathbf{\tilde{y}^{(k)}})^{T}\mathbf{\tilde{z}^{k+1}}\\
    &\implies (\mathbf{\tilde{y}^{(k+1)}} - \mathbf{\tilde{y}^{(k)}})^{T}(\mathbf{\tilde{z}^{(k+1)}} - \mathbf{\tilde{z}^{(k)}}) \geq 0.
\end{align*}
We substitute $\mathbf{\tilde{y}^{(k+1)}} - \mathbf{\tilde{y}^{(k)}} = \beta \mathbf{\tilde{r}^{(k+1)}}$ to get 
$$\beta (\mathbf{\tilde{r}^{(k+1)}})^{T}(\mathbf{\tilde{z}^{(k+1)}} - \mathbf{\tilde{z}^{(k)}}) \geq 0,$$
because $\beta > 0$ and subsequently
\begin{align*}
    V^{(k)} - V^{(k+1)} &\geq \beta||\mathbf{\tilde{r}^{(k+1)}} + (\mathbf{\tilde{z}^{(k+1)}} - \mathbf{\tilde{z}^{(k)}})||_{2}^{2} \\
    &\geq \beta||\mathbf{\tilde{r}^{(k+1)}}||_{2}^{2} + \beta||(\mathbf{\tilde{z}^{(k+1)}} - \mathbf{\tilde{z}^{(k)}})||_{2}^{2},
    \qed
\end{align*}
thus proving the third inequality. The convergence of ADMM is a consequence of the three inequalities \ref{ineq:1}, \ref{ineq:4}, and \ref{ineq:7}. Finally, the inequality \ref{ineq:7} can be rewritten as

$$V^{(k+1)} \leq V^{(k)} - \beta||\mathbf{\tilde{r}^{(k+1)}}||_{2}^{2} - \beta||(\mathbf{\tilde{z}^{(k+1)}} - \mathbf{\tilde{z}^{(k)}})||_{2}^{2},$$

which implies that $V^{(k)}$ decreases in each iteration, thus bounding $\mathbf{\tilde{y}^{(k)}}$ and $\mathbf{\tilde{z}^{(k)}}$ as $V^{(k)} < V^{0}$. In particular, note that inductively inequality \ref{ineq:7} leads to the following set of inequalities
\begin{align*}
    V^{(0)} &\geq V^{(1)} + \beta||\mathbf{\tilde{r}^{(1)}}||_{2}^{2} + \beta||(\mathbf{\tilde{z}^{(1)}} - \mathbf{\tilde{z}^{(0)}})||_{2}^{2},\\
    V^{(1)} &\geq V^{(2)} + \beta||\mathbf{\tilde{r}^{(2)}}||_{2}^{2} + \beta||(\mathbf{\tilde{z}^{(2)}} - \mathbf{\tilde{z}^{(1)}})||_{2}^{2},\\
    &\vdots\\
    V^{(k)} &\geq V^{(k+1)} + \beta||\mathbf{\tilde{r}^{(k+1)}}||_{2}^{2} + \beta||(\mathbf{\tilde{z}^{(k+1)}} - \mathbf{\tilde{z}^{(k)}})||_{2}^{2},\\
    &\vdots
\end{align*}
which, upon summation reveals
\begin{align*}
    \sum_{k=0}^{\infty} V^{(k)} &\geq \sum_{k=0}^{\infty} V^{(k+1)} + \beta \sum_{k=0}^{\infty} \Big(||\mathbf{\tilde{r}^{(k+1)}}||_{2}^{2} + ||(\mathbf{\tilde{z}^{(k+1)}} - \mathbf{\tilde{z}^{(k)}})||_{2}^{2}\Big)\\
    \sum_{k=0}^{\infty}( V^{(k)} - V^{(k+1)}) &\geq \beta \sum_{k=0}^{\infty} \Big(||\mathbf{\tilde{r}^{(k+1)}}||_{2}^{2} + ||(\mathbf{\tilde{z}^{(k+1)}} - \mathbf{\tilde{z}^{(k)}})||_{2}^{2}\Big)\\
    V^{(0)} &\geq \beta \sum_{k=0}^{\infty} \Big(||\mathbf{\tilde{r}^{(k+1)}}||_{2}^{2} + ||(\mathbf{\tilde{z}^{(k+1)}} - \mathbf{\tilde{z}^{(k)}})||_{2}^{2}\Big) \qquad\Big[ \sum_{k=0}^{\infty}( V^{(k)} - V^{(k+1)})\textrm{ is a telescopic series}\Big].
\end{align*}
Thus the sum of non-negative terms on the right-hand side  $\sum_{k=0}^{\infty}\Big(||\mathbf{\tilde{r}^{(k+1)}}||_{2}^{2} + ||(\mathbf{\tilde{z}^{(k+1)}} - \mathbf{\tilde{z}^{(k)}})||_{2}^{2}\Big) $
converges, as the $n^{th}$ partial sum is bounded above by  
$$\sum_{k=0}^{n}\Big(||\mathbf{\tilde{r}^{(k+1)}}||_{2}^{2} + ||(\mathbf{\tilde{z}^{(k+1)}} - \mathbf{\tilde{z}^{(k)}})||_{2}^{2}\Big) \leq \frac{V^{0}}{\beta}.$$
Thus, the primary residual, $\mathbf{\tilde{r}^{(k)}} \to 0$ and the secondary residual $\beta(\mathbf{\tilde{z}^{(k+1)}} - \mathbf{\tilde{z}^{(k)}}) \to 0$ as $k \to \infty$ immediately from $\beta >0$, completing the proof of Proposition 3.
\end{proof}

\section{Additional figures and tables}

\begin{figure}[H]
\begin{subfigure}{.50\textwidth}
  \centering
  \includegraphics[width=.8\linewidth]{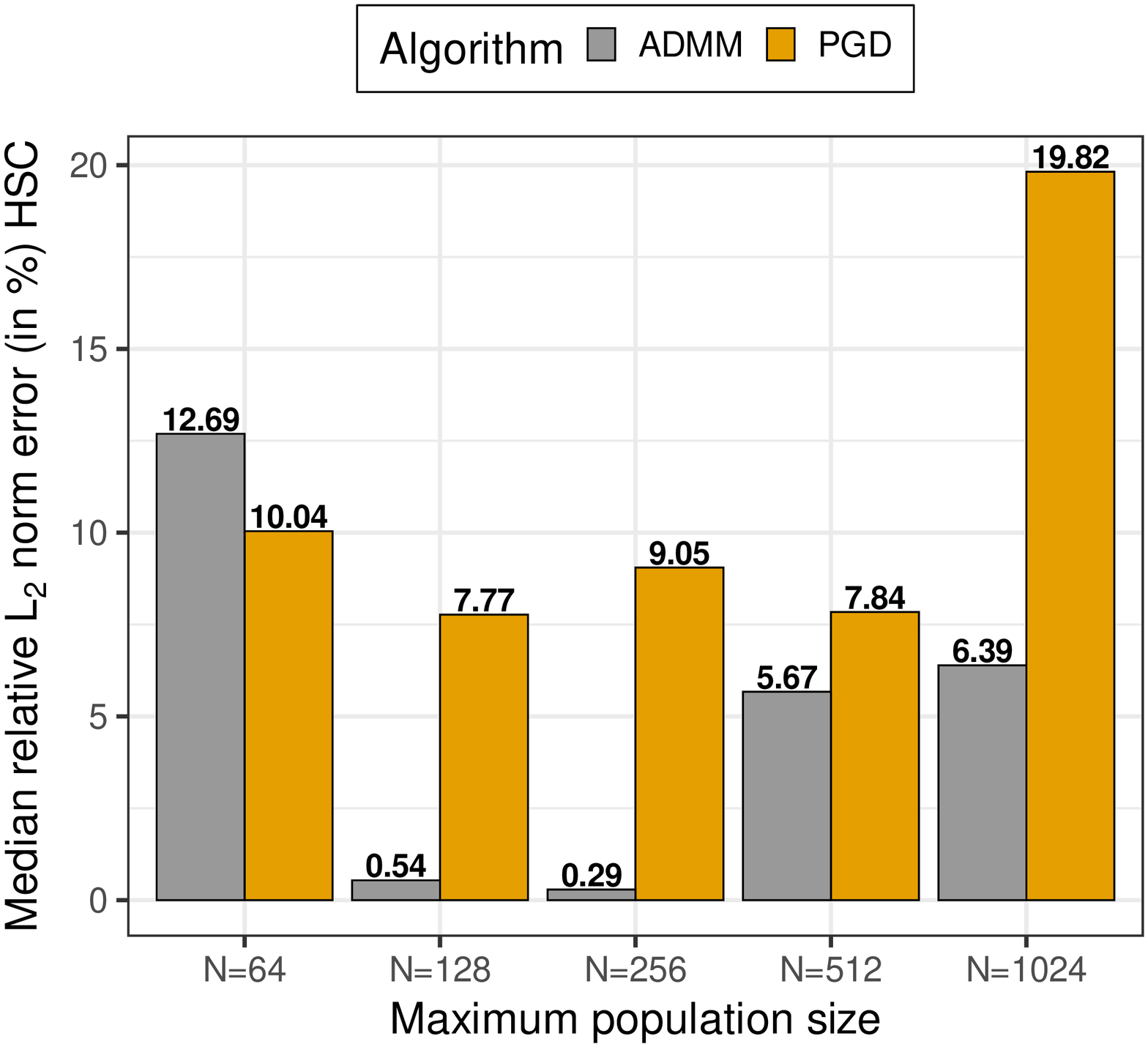}  
  \caption{}
  \label{fig:S3}
\end{subfigure}
\begin{subfigure}{.50\textwidth}
  \centering
  \includegraphics[width=.8\linewidth]{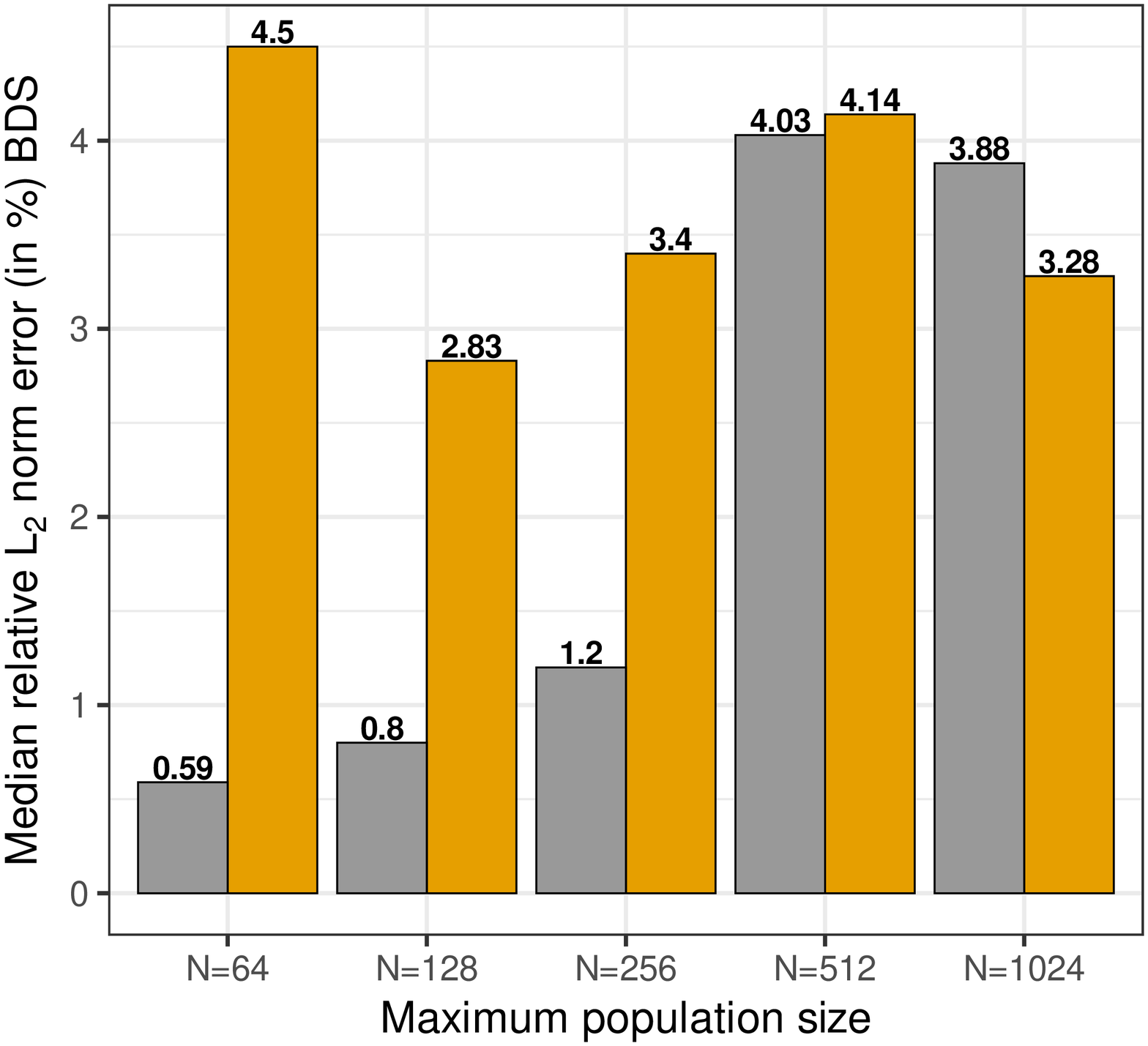}  
  \caption{}
  \label{fig:S4}
\end{subfigure}
\caption{The median relative norm errors $\epsilon^{L_{2}}_{rel}$ in recovering the transition probabilities using both PGD and ADMM algorithms for the a) HSC model and b) BDS model. Note that as opposed to the runtimes, the $\epsilon^{L_{2}}_{rel}$ was the same for both the vanilla and GPU implementations of the ADMM algorithm.}
\end{figure}

\begin{figure*}[htbp!]
    \centering
    \begin{subfigure}[t]{0.5\textwidth}
        \centering
        \includegraphics[height=1.2in]{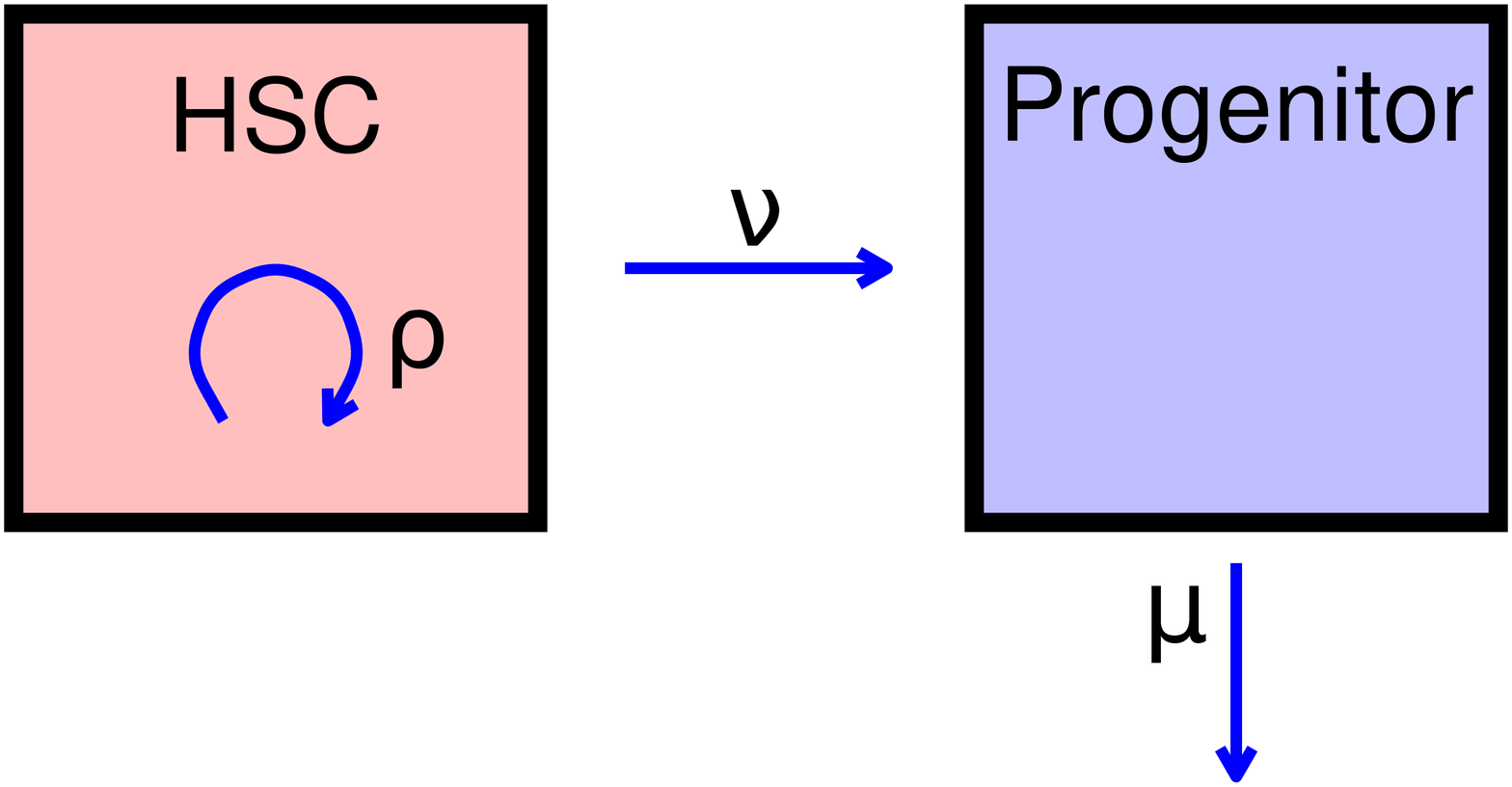}
        \caption{}
        \label{fig:S2}
    \end{subfigure}%
    ~ 
    \begin{subfigure}[t]{0.5\textwidth}
        \centering
        \includegraphics[height=2.0in]{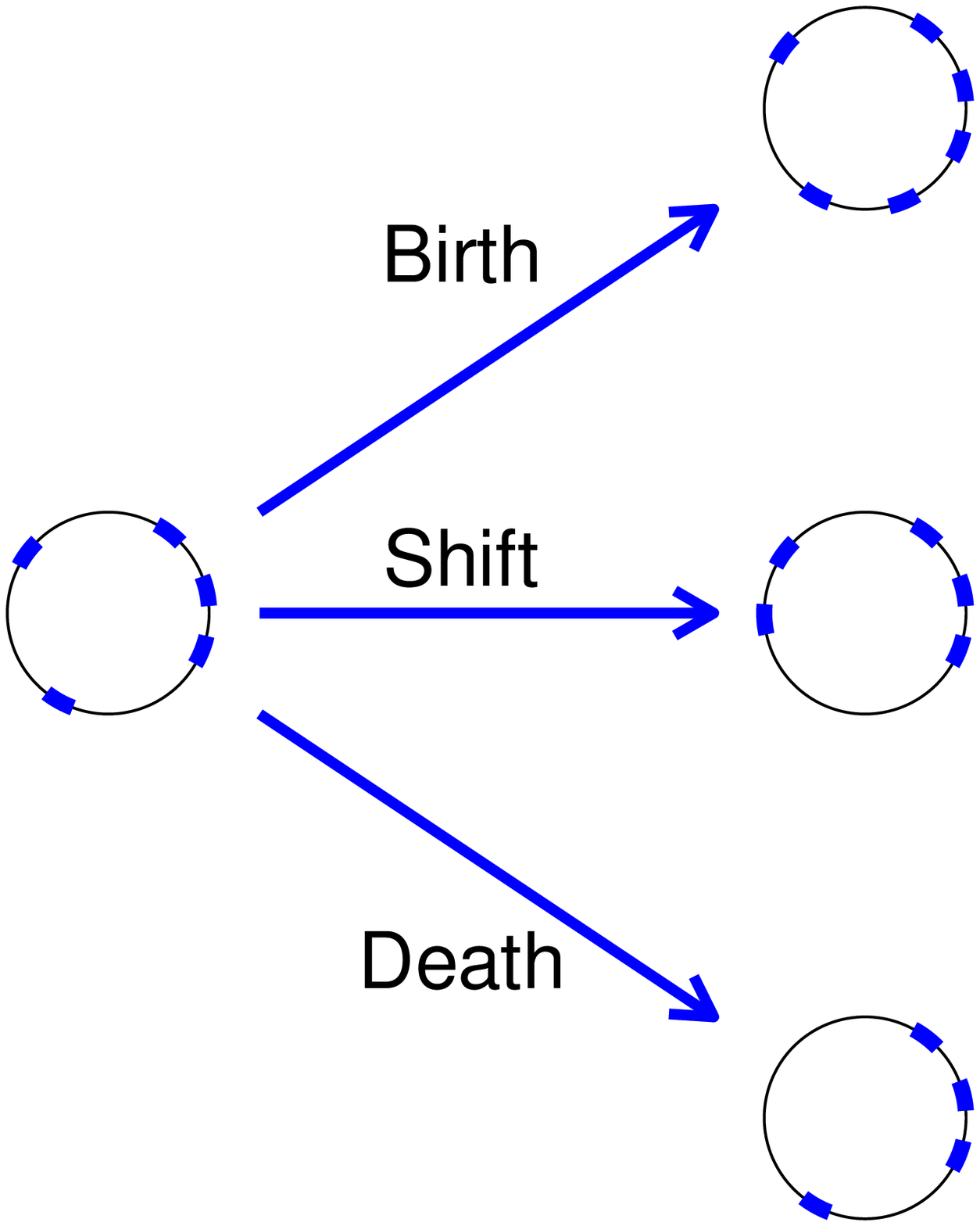}
        \caption{}
    \end{subfigure}
    \caption{a) HSCs can self-renew, producing new HSCs at rate $\rho$, or differentiate into progenitor cells at rate $\nu$. Further progenitor differentiation is modeled by rate $\mu$ b) Illustration of the three types of transposition - birth, death, shift - along a genome, represented by circles \citep{rosenberg2003estimating}. Transposons are depicted by rectangles occupying locations along the circles/genomes. On the right set of diagrams, a birth event keeps the number of type 1 particles intact and increments the number of type 2 particles by one, a death event changes the number of type 1 particles from five to four and keeps the number of type 2 particles at zero, and finally a shift event decreases the number of type 1 particles by one and increases the number of type 2 particles by one.}
\end{figure*}

\begin{table}[ht]
\centering
\begin{tabular}{clllll|lllll}
\hline
 & \multicolumn{1}{c}{} & \multicolumn{3}{c}{\textbf{HSC Model}} & \multicolumn{1}{c|}{} & \multicolumn{3}{c}{\textbf{BDS Model}} & \multicolumn{1}{c}{} & \multicolumn{1}{c}{} \\ \hline
 & \multicolumn{1}{c}{$N\!=\!64$} & \multicolumn{1}{c}{$N\!=\!128$} & \multicolumn{1}{c}{$N\!=\!256$} & \multicolumn{1}{c}{$N\!=\!512$} & \multicolumn{1}{c|}{$N\!=\!1024$} & \multicolumn{1}{c}{$N\!=\!64$} & \multicolumn{1}{c}{$N\!=\!128$} & \multicolumn{1}{c}{$N\!=\!256$} & \multicolumn{1}{c}{$N\!=\!512$} & \multicolumn{1}{c}{$N\!=\!1024$} \\ \hline
Stepsize $(\beta$) & $0.08$ & $0.005$ & $0.08$ & $0.005$ & $0.005$ & $0.005$ & $0.005$ &  $0.005$ & $0.0005$ & $0.0005$ \\ \hline
$D_{1}$ & $N^{2}$ & $N^{2}$ & $N^{2}$ & $N^{2}$ & $N^{2}$ & $N^{2}$ & $N^{2}$ & $N^{2}$ & $N^{2}$ & $N$  \\ \hline
$D_{2}$ & $N^{5}$ & $N^{5}$ & $N^{5}$ & $N^{5}$ & $N^{5}$ & $N^{2}$ & $N^{2}$ & $N^{2}$ & $N^{2}$ & $N$ \\ \hline
$\epsilon_{abs}$ & $10^{-2}$ & $10^{-2}$ & $10^{-2}$ & $10^{-2}$ & $10^{-3}$ &  $10^{-3}$ & $10^{-3}$ & $10^{-3}$ &$10^{-3}$ & $10^{-3}$  \\ \hline
$\epsilon_{abs}$ & $10^{-3}$ & $10^{-3}$ & $10^{-3}$ &$10^{-3}$ & $10^{-3}$ &  $10^{-3}$ & $10^{-3}$ & $10^{-3}$ &$10^{-3}$ & $10^{-3}$ \\ \hline
\end{tabular}
\vspace{2mm}
\label{tab:S1}
\caption{This table provides complete details of the ADMM algorithm under our proposed framework, including the step size $\beta$, the tolerances for the primal and dual feasibility conditions, $\epsilon_{pri} = D_{1} \epsilon_{abs} + \epsilon_{rel} \max(||\pmb{u^{(k)}}||_{2}, ||\pmb{z^{(k)}}||_{2})$ and $\epsilon_{dual} = D_{2} \epsilon_{abs} + \epsilon_{rel}, ||\pmb{y^{(k)}}||_{2}$ respectively, and $\epsilon_{abs}$ and 
$\epsilon_{rel}$ are the absolute and relative tolerances respectively. The regularization parameter $\lambda$ was chosen to be $0.5(\log M)$ for both models. These were chosen in a simple manner across \textit{all} trials but our framework still yielded promising performance results as compared to the PGD algorithm.}
\end{table}

Tables \ref{tab:S2} and \ref{tab:S3} show the median running times for different algorithms namely, PGD, ADMM (Vanilla), and ADMM (GPU) and the percentage change in runtimes relative to the runtime of the PGD algorithm,
$$ \text{\% change} = \frac{\text{runtime}_{ADMM} - \text{runtime}_{PGD}}{\text{runtime}_{PGD}},$$
for the HSC and BDS models, respectively. The negative percent change indicates that the ADMM algorithm is faster than the PGD algorithm for the corresponding $N$ value. 

\begin{table}[htbp!]
\centering
\begin{tabular}{lllll}
& & \textbf{Median runtime} & \\ \hline
\textbf{Maximum population size} & {M} & \textbf{PGD} & \textbf{ADMM (\% change)} & \textbf{ADMM (GPU) (\% change)}\\ \hline
$N\!=\!64$ & $51$ & $0.179$ & $0.009$ $(-94.97\%)$ & $0.016$ $(-91.06\%)$ \\ \hline
$N\!=\!128$ & $78$ & $0.911$ & $0.376$ $(-58.73\%)$ & $0.141$ $(-84.52\%)$ \\ \hline
$N\!=\!256$ & $83$ & $4.662$ & $1.951$ $(-58.15\%)$ & $0.176$ $(-96.22\%)$ \\ \hline
$N\!=\!512$ & $88$ & $29.34$ & $4.721$ $(-83.91\%)$ & $0.227$ $(-99.23\%)$ \\ \hline
$N\!=\!1024$ & $93$ & $869.49$ & $20.44$ $(-97.65\%)$ & $0.983$ $(-99.89\%)$ \\ \hline
\end{tabular}
\vspace{2mm}
\caption{A comparison of median running times (percent change relative to PGD algorithm runtime) for different algorithms and scales for the HSC model. For all maximum population sizes, our proposed framework is faster in recovering the transition probabilities than the PGD algorithm for errors at least as good as PGD.}
\label{tab:S2}
\end{table}

\begin{table}[htbp!]
\centering
\begin{tabular}{lllll}
& & \textbf{Median runtime} & \\ \hline
\textbf{Maximum population size} & {M} & \textbf{PGD} & \textbf{ADMM (\% change)} & \textbf{ADMM (GPU) (\% change)}\\ \hline
$N\!=\!64$ & $18$ & $0.495$ & $0.195$ $(-60.60\%)$ & $0.104$ $(-78.99\%)$ \\ \hline
$N\!=\!128$ & $19$ & $1.50$ & $1.09$ $(-27.33\%)$ & $0.196$ $(-86.93\%)$ \\ \hline
$N\!=\!256$ & $29$ & $13.02$ & $3.50$ $(-86.93\%)$ & $0.216$ $(-98.34\%)$ \\ \hline
$N\!=\!512$ & $22$ & $113.04$ & $20.23$ $(-82.10\%)$ & $0.524$ $(-99.54\%)$ \\ \hline
$N\!=\!1024$ & $28$ & $459.93$ & $132.85$ $(-71.12\%)$ & $3.303$ $(-99.28\%)$ \\ \hline
\end{tabular}
\vspace{2mm}
\caption{A comparison of median running times (percent change relative to the runtime of the PGD algorithm) for different algorithms and scales for the BDS model. For all maximum population sizes, our proposed framework is faster in recovering the transition probabilities than the PGD algorithm for errors at least as good as PGD.}
\label{tab:S3}
\end{table}

\section{Details on Generating Functions and PGD Algorithm}

\subsection{Pseudocode for PGD}

The following pseudocode summarizes the proximal gradient descent approach proposed in \citet{xu2015efficient}.
\begin{algorithm}[H]
\caption{PGD Algorithm} 

    \SetKwInOut{Input}{Input}
    \SetKwInOut{Output}{Output}

\underline{function PGD} $(B,\beta, \tau, \gamma, \epsilon_{abs}, \epsilon_{rel}, N_{iter}, P)$\;
    \Input{initial sizes $X_{1} = j, X_{2} = k$, time interval $t$, branching rates $\theta$, signal size $N > j,k$, measurement size $M$, penalization constant $\lambda>0$, line-search parameters $L,c$.}
    \Output{A real-valued 2D matrix $\pmb{\hat{S}}$}
      Uniformly sample $M$ indices $\mathcal{J} \subset [0, \ldots N]$\\
      Compute $\pmb{b} = \{\phi_{j,k}(t,e^{2\pi iu/N}, e^{2\pi iv/N} )\}_{u,v \in \mathcal{J}}$\\
      Define \textbf{A}$ = \psi_{\mathcal{J}}$, the $\mathcal{J}$ rows of the IDFT matrix $\psi$.\\
      \textbf{Initialize: $\pmb{S_{1} = Y_{1} = 0}$}\\
      \For {$k=1,2,\ldots$}{
      Choose $L_{k}$ = line-search$(L,c,Y_{k})$\\
      Update extrapolation parameter: $\omega_{k} = \frac{k}{k+3}$\\
      Update momentum: $\pmb{Y_{k+1} = S_{k} + \omega_{k}(S_{k} - S_{k-1})}$\\
      Update: $S_{k+1} = softh(S_{k} - L_{k})$ 
      }
      \textbf{return} $\hat{S} = S_{k+1}$
    \label{alg:2}  

\end{algorithm}

\subsection{Derivation of the PGF for HSC model}
The details for deriving the PGF are included here only for completeness, but are standard, following the ``random variable technique" of \citep{bailey1991elements}. Given a two-type branching process with instantaneous rates $a_{i}(k,l)$, we can define the pseudo-generating function for $i=1,2$,
$$u_{i}(s_{1},s_{2}) = \sum_{k} \sum_{l} a_{i}(k,l)s_{1}^{k}s_{2}^{l}.$$

The probability generating function can be expanded as
\begin{align*}
    \phi_{10}(t,s_{1}, s_{2}) &= \mathbb{E}(s_{1}^{X_{1}(t)}s_{2}^{X_{2}(t)}|X_{1}(0)=1, X_{2}(0) = 0)\\
    &= \sum_{k} \sum_{l} p_{(1,0),(k,l)}(t)s_{1}^{k}s_{2}^{l}\\
    &= \sum_{k} \sum_{l} (\delta_{k=1,l=0}+a_{1}(k,l)t +o(t))s_{1}^{k}s_{2}^{l}\\
    &= s_{1} + u_{1}(s_{1},s_{2})t + o(t).
\end{align*}

Similarly, starting with one particle of type $2$ instead of type 1, we can obtain an analogous expression for $\phi_{01}(t,s_{1}, s_{2})$. For brevity, we will write $\phi_{10}(t,s_{1}, s_{2}) := \phi_{1}(t,s_{1}, s_{2}), \phi_{01}(t,s_{1}, s_{2}) := \phi_{2}(t,s_{1}, s_{2})$. \\

Differentiating $\phi_{1}(t,s_{1}, s_{2})$ and $\phi_{2}(t,s_{1}, s_{2})$ with respect to time, we get
\begin{align*}
    \frac{d \phi_{1}}{dt}\big(t,s_{1},s_{2}\big)|_{t=0} &= u_{1}(s_{1},s_{2})\\
    \frac{d \phi_{2}}{dt}\big(t,s_{1},s_{2}\big)|_{t=0} &= u_{2}(s_{1},s_{2}).
\end{align*}
We now derive the backward and forward equations with the Chapman-Kolmogorov equations, yielding the following symmetric relations:
\begin{align}
    \phi_{1}(t+h,s_{1},s_{2}) &= \phi_{1}(t,\phi_{1}(h,s_{1},s_{2}), \phi_{2}(h,s_{1},s_{2})) \label{eq:24}\\
    &= \phi_{1}(h,\phi_{1}(t,s_{1},s_{2}), \phi_{2}(t,s_{1},s_{2})).
\end{align}
We expand around $t$ and apply Equation \ref{eq:24} to derive the backward equations:

\begin{align*}
    \phi_{1}(t+h,s_{1},s_{2}) &= \phi_{1}(t,s_{1},s_{2}) + \frac{d \phi_{1}}{dh}\big(t+h,s_{1},s_{2}\big)|_{h=0}h + o(h)\\
    &= \phi_{1}(t,s_{1},s_{2}) + \frac{d \phi_{1}}{dh}\big(h,\phi_{1}(t,s_{1},s_{2}), \phi_{2}(t,s_{1},s_{2})\big)|_{h=0}h + o(h) \\
    &= \phi_{1}(t,s_{1},s_{2}) + u_{1}(\phi_{1}(t,s_{1},s_{2}), \phi_{2}(t,s_{1},s_{2})) + o(h)
\end{align*}

We can apply an analogous argument for $\phi_{2}(t,s_{1},s_{2})$ to arrive at the following system of ODEs
\begin{align*}
    \frac{d \phi_{1}}{dt}\big(t,s_{1},s_{2}\big) &= u_{1}(\phi_{1}(t,s_{1},s_{2}) + \phi_{2}(t,s_{1},s_{2})) \\
    \frac{d \phi_{2}}{dt}\big(t,s_{1},s_{2}\big) &= u_{2}(\phi_{1}(t,s_{1},s_{2}) + \phi_{2}(t,s_{1},s_{2}))
\end{align*}
with initial conditions $\phi_{1}(0,s_{1},s_{2})=s_{1}$, $\phi_{2}(0,s_{1},s_{2}) = s_{2}$. 

Recall from the main text the rates defining the two-component HSC model are given by:
$$a_{1}(2,0) = \rho, \quad a_{1}(0,1) = \nu, \quad a_{1}(1,0) = -(\rho+\nu), \quad a_{2}(0,0) = \mu, \quad a_{2}(0,1) = -\mu.$$

Thus, the pseudo-generating functions become
\begin{align*}
    u_{1}(s_{1},s_{2}) =& \rho s_{1}^{2} + \nu s_{2} -(\rho+\nu)s_{1}, \\
     u_{2}(s_{1},s_{2}) =& \mu - \mu s_{2} = \mu(1-s_{2}).
\end{align*}
Plugging these into the backward equations yields,
\begin{align*}
    \frac{d \phi_{1}}{dt}\big(t,s_{1},s_{2}\big) &= \rho\phi_{1}^{2}(t,s_{1},s_{2}) + \nu \phi_{2}(t,s_{1},s_{2}) -(\rho+\nu) \phi_{1}(t,s_{1},s_{2}), \\
    \frac{d \phi_{2}}{dt}\big(t,s_{1},s_{2}\big) &= \mu(1 - \phi_{2}(t,s_{1},s_{2})).
\end{align*}
Notice that the differential equation for $\phi_{2}(t,s_{1},s_{2})$ corresponds to that of a pure death process and has a closed-form solution.

\begin{align*}
    \frac{d \phi_{2}}{dt}\big(t,s_{1},s_{2}\big) &= \mu(1 - \phi_{2}(t,s_{1},s_{2}))\\
    \frac{d}{dt}\Big(\frac{\phi_{2}}{1-\phi_{2}}\Big)\big(t,s_{1},s_{2}\big) &= \mu\\
    \phi_{2}(t,s_{1},s_{2}) &= 1 - e^{-\mu t + C}
\end{align*}
Plugging in $\phi_{2}(0,s_{1},s_{2}) = s_{2}$, we obtain $C=ln(1-s_{2})$, and we get
\begin{equation}
    \phi_{2}(t,s_{1},s_{2}) = 1 + (s_{2}-1)e^{-\mu t}.
    \label{eq:25}
\end{equation}

On plugging Equation \ref{eq:25} into the backward equation involving $\phi_{1}(t,s_{1},s_{2})$, we obtain
\begin{equation}
    \frac{d \phi_{1}}{dt}\big(t,s_{1},s_{2}\big) = \rho\phi_{1}^{2}(t,s_{1},s_{2}) -(\rho+\nu) \phi_{1}(t,s_{1},s_{2}) + \nu (1 + (s_{2}-1)e^{-\mu t})
    \label{eq:26}
\end{equation}
Given the rates of the process and the values for the three arguments, Equation \ref{eq:26} can be solved numerically, allowing the computation of $\phi_{i,j}(t,s_{1}, s_{2}) = \phi_{1}^{i}(t,s_{1}, s_{2})\phi_{2}^{j}(t,s_{1}, s_{2})$  which holds by particle independence. 

\end{appendices}
\end{document}